\documentclass[a4paper,twoside,10pt]{letter}
\usepackage{graphicx,saj,multicol,subeqnarray}

\usepackage{amsmath,colortbl}

\usepackage{pdflscape}
\usepackage{afterpage}
\usepackage{capt-of} 
\usepackage{subcaption}

\def\papertype{Editorial}

\def\udc{...}
\begin{document}
\baselineskip=3.1truemm
\columnsep=.5truecm
\newenvironment{lefteqnarray}{\arraycolsep=0pt\begin{eqnarray}}
{\end{eqnarray}\protect\aftergroup\ignorespaces}
\newenvironment{lefteqnarray*}{\arraycolsep=0pt\begin{eqnarray*}}
{\end{eqnarray*}\protect\aftergroup\ignorespaces}
\newenvironment{leftsubeqnarray}{\arraycolsep=0pt\begin{subeqnarray}}
{\end{subeqnarray}\protect\aftergroup\ignorespaces}
%


\markboth{\eightrm ACs discovered in a sample of Galactic short period TIICs} {\eightrm M. I. JURKOVIC}

{\ }

\publ

\type

{\ }


\title{Anomalous Cepheids discovered in a sample of Galactic short period Type II Cepheids}


\authors{M. I. Jurkovic$^{1,2}$}

\vskip3mm

\address{$^1$Astronomical Observatory, Volgina 7, 11060 Belgrade, Serbia}

\Email{mojur}{aob.rs}

\address{$^2$Konkoly Thege Mikl\'{o}s Astronomical Institute, Research Centre for Astronomy and Earth Sciences, Hungarian Academy of Sciences\break H-1121 Budapest, Konkoly Thege Mikl\'{o}s \'{u}t 15-17.}


\dates{xxxx xx, 2018}{xxxx xx, 2018}


\summary{We revisited the short period Type II Cepheids (T2Cs), called the BL Herculis (BLHs), in the Galactic Field to derive a homogeneous analysis of their Fourier parameters.
\noindent Only V-band data were compiled to make sure that it was directly comparable between the known variables of the OGLE-III catalogue and the 59 individual objects classified as short period Type II Cepheids in the General Catalogue of Variable Stars (GCVS) we had in our sample. The derived Fourier parameters were used to make the distinction between different classes of variables.
\noindent From the 59 stars we found 19 BLHs, 19 fundamental mode Anomalous Cepheids (ACs) (8 of them were already known from the Catalina Sky Survey (CSS)), 1 first overtone AC, 2 were found to be possible peculiar W Virginis (pWVir), 11 classical Cepheids (DCEPs), and 7 stars were not pulsating variables at all. As a result we created a list of bright BLH stars in the Galactic Field, and separated the ACs, as well as other objects that were misclassified. The number of true BLHs decreased in our sample by more than 50\%. We gathered the metallicity from spectroscopic measurements published in the literature. While the number of actual measurements is low, it is highly suggestive that ACs are metal poor. The mean metallicity from 8 measurements in 4 stars (UY Eri having 5 different [Fe/H] data points) is -1.12 dex, but if the higher value metallicity outliers of UY Eri are left out the mean metallicity becomes -1.88 dex, regardless if the AC is in the Milky Way itself or in a cluster. On the other hand, BLHs seem to have a Solar-like metallicity of 0 dex averaged from 21 measurements of 10 stars.}



\begin{multicols}{2}


\section{1. INTRODUCTION}
\label{sec:intro}

	The short period Type II Cepheids (T2Cs), called BL Herculis (BLH) subtype, are low mass ($\approx$ 0.5 - 0.6 M$_\odot$) pulsating stars, with periods between 1 - 4 days. BLHs pulsate only in the fundamental mode (F). A summary of these objects was given in Wallerstein (2002) and Catelan \& Smith (2015). In the past the General Catalogue of Variable Stars (GCVS\footnote{http://www.sai.msu.su/gcvs/gcvs/}, Samus et al. 2009a, Samus et al. 2018) was the biggest source of the classification of variable stars, but with the emergence of the big sky observing programs their lists became somewhat outdated, so we decided to revisit the BLHs subtype to see how they compare with the other datasets. In the GCVS the BLHs are marked with "CWB" and candidates with "CWB:". The boarder between the short period and longer period W Virginis (CWA) stars was put at 8 days, which got outdated by the results of the T2Cs from the Optical Gravitational Lensing Experiment III (OGLE-III) catalogue  (Soszy\'{n}ski et al. 2008, 2010b, 2011b). In the latest OGLE-IV catalogue composed of stars in the Galactic bulge (Soszy\'{n}ski et al. 2017) the border between BLHs and WVir stars was moved from 4 to 5 days for the Bulge.

	Anomalous Cepheids (ACs) are stars which have similar period range ($0.24 - 4$ days), but their masses are higher 1.2-1.8 M$_{\odot}$. In the instability strip (IS) of the Herztsprung-Russell diagram (HRD) they are just above the T2Cs, see Bono et al. (1997b), Fiorentino et al. (2006), Groenewegen and Jurkovic (2017b). ACs pulsate in the fundamental mode (F) and in the first overtone (1O). In the OGLE-III catalogue we have seen  (Soszy\'{n}ski et al. 2008, 2010b,a) that T2Cs and ACs form a different Period-Luminosity ($PL$) relations in the LMC and SMC. Groenewegen and Jurkovic (2017a,b) confirmed this in their papers calculating the luminosities ($L$), effective temperatures ($T_{\rm eff}$), masses ($M$), radii ($R$) and the bolometric $PL$ relation for the T2Cs and ACs in the LMC and SMC. The number ACs identified in the LMC and SMC in the OGLE-IV catalogue is 174 F and 76 1O pulsators. In Soszy\'{n}ski et al. (2017) the number of identified ACs in the Galactic bulge has risen to 20 (19 F and 1 1O). 
	
	In Catelan and Smith (2015) it is stated that most of ACs have been discovered in dwarf spheroidal galaxies, in the Magellanic Clouds (MCs), and a few in globular clusters, but only a handful in the Milky Way (MW). The whole subgroup of ACs is also know as BLBOO in the GCVS, and it was named after the variable BL Bootis, which was considered to be the only AC in the MW, and it turned out to be a member of the galactic globular cluster NGC~5466. The discovery was made by Zinn and Dahn (1976), and studied in detail e.g. by McCarthy and Nemec (1997), Nemec and McCarthy (1998). Szabados et al. (2007) published a detailed study of XZ Ceti, confirming it to be a 1O AC, making it the second AC discovered in the MW. 

	Articles in the literature about T2Cs might further complicate the understanding of the classification, since the derived properties of these variables were interpreted differently. We give a short overview of these papers, but a longer summary is given in Catelan and Smith (2015). Diethelm (1983) has published a classification of pulsating stars based on the shape of their light curves. V716 Oph, BF Ser, CE Her, VX Cap, XX Vir, EK Del, UY Eri and UX Nor were labelled RR Lyrae Type \textit{d} (RRd). Petersen and Diethelm (1986) concluded that the properties of the T2Cs known at that time is not uniform, and the stars  known as RRds are singled out as forming a different group. To make things more complicated the nomenclature of the RR Lyrae (RRL) variables has added a subclass also called RRd, which describes an RRL simultaneously pulsating in the F and 1O, with a period ratio of $0.745$, and the upper limit of the RRL variables is established to be at 1 day (e.g. see Catelan and Smith (2015)).
	
	The Catalina Sky Survey (CSS), see Drake et al. (2014a), published their catalogue and they have detected 64 ACs in total Drake et al. (2014b). The following six stars overlap with our BLH sample: FY Vir, BF Ser, VX Cap, XX Vir, V1149 Her, and V716 Oph. Then in 2017 they published a new Catalina Southern Survey (Drake et al. 2017) with additional 153 ACs, and two more stars overlapped with our sample: V563 Cen and BI Tel. Most recently   Soszy\'{n}ski et al. (2017) added 20 new ACs detected in toward the Galactic center by the OGLE-IV program.
	
	Wallerstein (2002) dealt with the question of metallicities of T2Cs. Most of the data available at that point were estimated from different photometric systems (Diethelm 1990, Harris 1981). Nevertheless, there was a  separation between a group of stars that were very metal-poor, and the once that had metallicities larger than $-$0.3.

	We have looked into 128 stars in the GCVS, but found V-band data for only 59 stars that were previously classified in the GCVS as CWB stars. Applying the Fourier analysis, and comparing our results to the V-band dataset of ACs and T2Cs in the LMC from OGLE-III we reclassified the above mentioned 59 stars comparing their Fourier parameters to the OGLE-III sample. We have made clear which of these objects are known cluster members (open and globular), since their evolution can be linked to the history of the evolution of the clusters that is separate from the individual stars in the Galaxy. The spectroscopically measured metallicites, as well as, other derived physical parameters were  gathered from the literature. These two variable types are important to be separated, since they do have significantly different ages, as a result of having different initial masses, and the clearance on their true numbers in the MW should be useful to future research in variety of fields, for example in the evolutionary modelling of stars or understanding the star formation history of our Galaxy. T2Cs are old objects making it possible to constrain some early star forming regions to a specific metallicity content helping future galactic evolution modelling. Because the origin of ACs is still not clear - they could be either a result of interacting binaries or they could have come from single star evolution - their contribution to galactic archaeology is not that straightforward, but because they are found in dwarf spheroidal and irregular galaxies as well as spiral galaxies (such as the MW or the LMC) and sometimes in globular cluster and they are metal-poor they could give an insight to the intermediate-age (1-6 Gyrs) star formation mechanism in these systems.
	
\section{2. THE DATA}
\label{sec:data}

	The comparison sample was taken from the OGLE-III catalog for the LMC BLHs and ACs. The Fourier parameters published in the online catalogue are derived from the I-band data, so we have collected the OGLE-III Soszy\'{n}ski et al. (2008, 2010b) V-band data and did the Fourier analysis ourselves. 

	In the Milky Way we have stared with the 128 star classified in the GCVS as short period T2Cs or possible T2Cs. We were able to find datasets for 59 stars in that list. We have collected the available data from the All Sky Automated Survey (ASAS-3\footnote{http://www.astrouw.edu.pl/asas/}),see Pojmanski (1997), in V-band for 34 stars. This search did not include all the objects in the GCVS list, so additional light curve data were collected from various publicly available databases: CSS\footnote{http://nesssi.cacr.caltech.edu/DataRelease/} (Drake et al. 2009, 2014), INTEGRAL\footnote{https://sdc.cab.inta-csic.es/omc/} and individual published papers, such as Berdnikov (2008), Henden (1980), Kwee and Diethelm (1984), Schmidt and Reiswig (1993), Schmidt et al. (2005), Soszy\'{n}ski et al. (2011b), and in one case data from the American Association of Variable Star Observers (AAVSO\footnote{https://www.aavso.org/}). We have added the source of the data for each star in the Appendix Table A.1. 
	
	While there are other sources (for example the SuperWASP\footnote{http://www.superwasp.org/}), with very good data, we could not use it for our comparative Fourier analysis, because they were collected in a broad band or no filter. 

\section{3. THE FOURIER ANALYSIS AND CLASSIFICATION PROCEDURE}
\label{sec:Fourier}

	We have used the Period04 program (Lenz and Breger 2004), which uses this equation (Eq.~\ref{eq:Fourier}):
\begin{equation}
    A = Z + \sum A_i sin(2\Pi(\omega_i t + \phi_i)).
	\label{eq:Fourier}
\end{equation}	

	 We have proceeded to get the Fourier parameters, defined by these equations:

\begin{equation}
    R_{i1}=\frac{A_i}{A_1},
	\label{eq:R_i1}
\end{equation}

and 

\begin{equation}
   \phi_{i1}=\phi_i - i\phi_1,
	\label{eq:phi_ik}
\end{equation}

where $i=2,3$, for the light curve analysis.

	This kind of Fourier decomposition was presented by Petersen and Diethelm (1986) based on the amplitude ratio ($R_{i1}$) and phase differences ($\phi_{i1}$) defined in Simon and Lee (1981). It is a useful tool for the automation of the classification. 

	With the publication of the OGLE-III data, a long spanning precise photometric information became available, making it possible to compare other stars to the known variables of the OGLE sample. We have concentrated our attention on the LMC dataset, because the Wessenheit index vs. log P for T2Cs and the F and 1O ACs separate very clearly (this is true for the SMC too, but we omitted that sample, because the number of AC was much smaller than in the LMC). When we go back to the Fourier parameters, we can be assured that this is a good indication of the type, and this is the sample we have compared our 59 Field stars against. All the Fourier parameters were converted to match the once from the OGLE-III catalogue.

	The calculated values of the Fourier parameters are given in Appendix, see Tables A.1, A.2 and A.3. The Fourier parameters are plotted in Figs. 1a, 1b, 1c, 1d. The ACs from the OGLE-III LMC sample and from the MW are plotted with purple, while the BLH from both datasets are green, to help the reader distinguish between the types of the Fourier parameter plots. The DCEPs are dark grey, and the possible pWVirs stars are orange.

	In addition to the Fourier parameters we have looked into the amplitudes of the detected harmonics ($A_2$ and $A_3$). Figs. 2 and 3 show how the T2Cs and ACs separate for the LMC and Galactic stars (the color coding is the same as in Figs. 1a, 1b, 1c and 1d). ACs have higher harmonic frequency amplitudes than T2Cs. There are a few stars that are overlapping. In the case of BQ CrA and V745 Oph this can be a consequence of the noisy data set that was used for the Fourier analysis. In the case of UY Eri the discrepancy in the Figs. 2 and 3 is attributable to the behaviour of the star. 

	The classification is based on an iterative process. Since the the OGLE BLHs and ACs are separated on their period-luminosity relation, we took their Fourier parameters as a comparison sample. To make sure that all the Fourier parameters are homogeneous we have used only V-band data for each examined star. The first step was to see where the OGLE LMC and MW stars overlap on all the six Fourier parameter plots. The $\phi_{21}$ parameter was the least helpful in distinguishing these types. This method,, however, does not give a clear classification for each object, since the Fourier parameters overlap in some cases. For this reason we did not apply any statistical criteria for the classification. Instead, when the classification of an object was not clear we looked at the light curve shape visually to confirm the result, and furthermore reviewed the previously published literature for individual stars to verify the result. The BLHs and ACs separated most clearly on the Figs. 2 and 3. In the cases where we felt that a star showed some peculiarities we have expanded our finding in individual paragraph.


\end{multicols}

\begin{minipage}[b]{.4\linewidth}
\centering
\includegraphics[width=0.9\columnwidth, keepaspectratio]{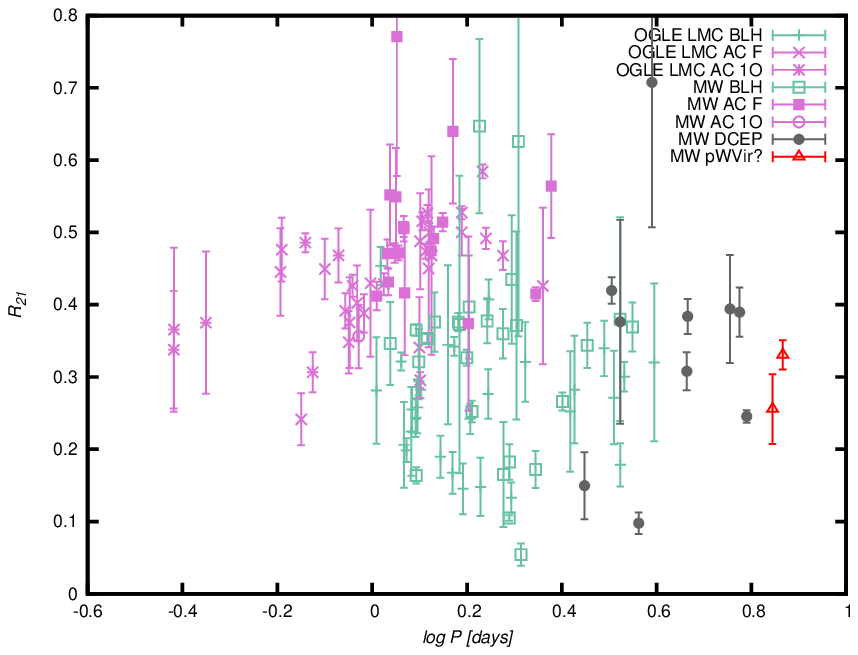}
\\
{(a)} 
\end{minipage}
\begin{minipage}[b]{.4\linewidth}
\centering
\includegraphics[width=0.9\columnwidth, keepaspectratio]{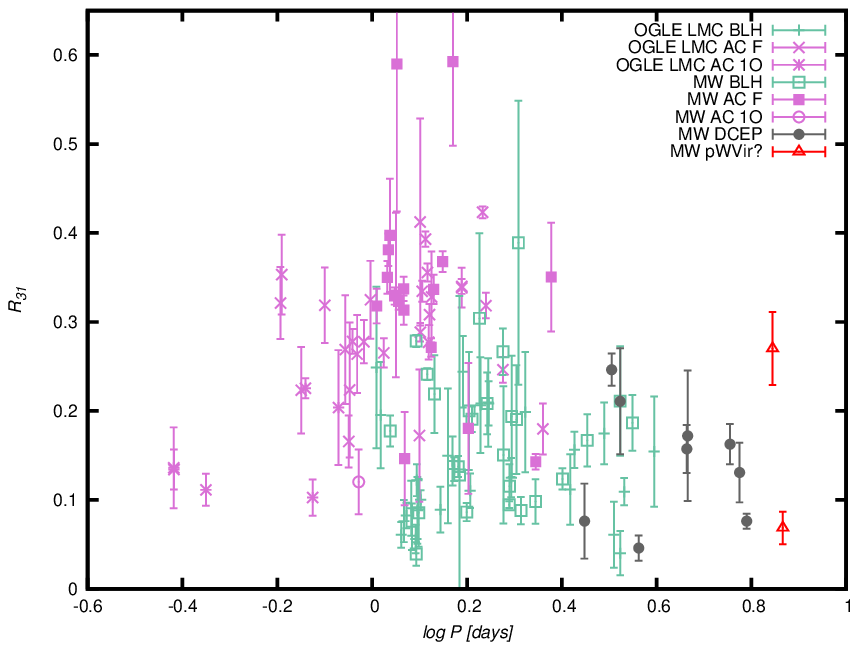}
\\
{(b)} 
\end{minipage}

\begin{minipage}[b]{.4\linewidth}
\centering
\includegraphics[width=0.9\columnwidth, keepaspectratio]{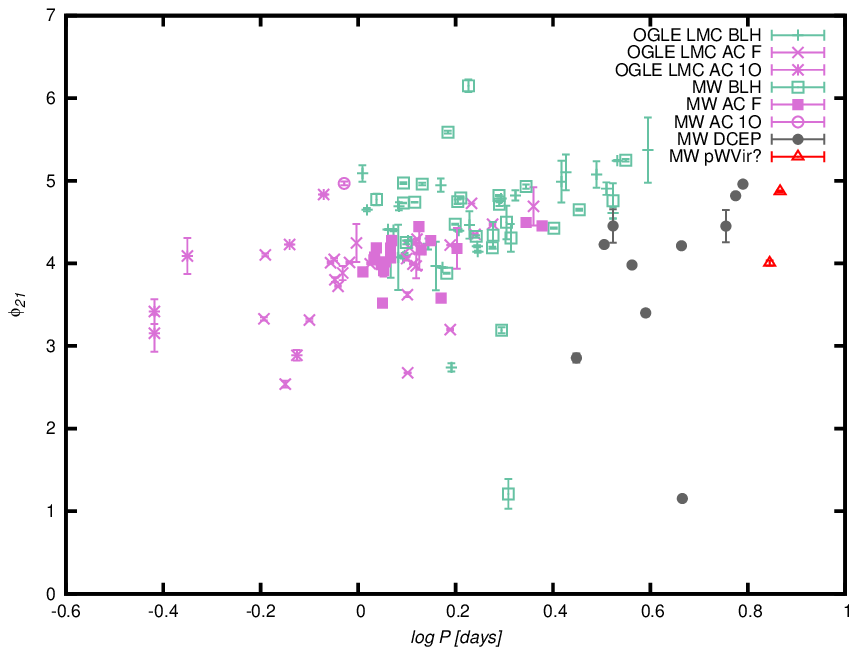}
\\
{(c)}
\end{minipage}
\begin{minipage}[b]{.4\linewidth}
\centering
\includegraphics[width=0.9\columnwidth, keepaspectratio]{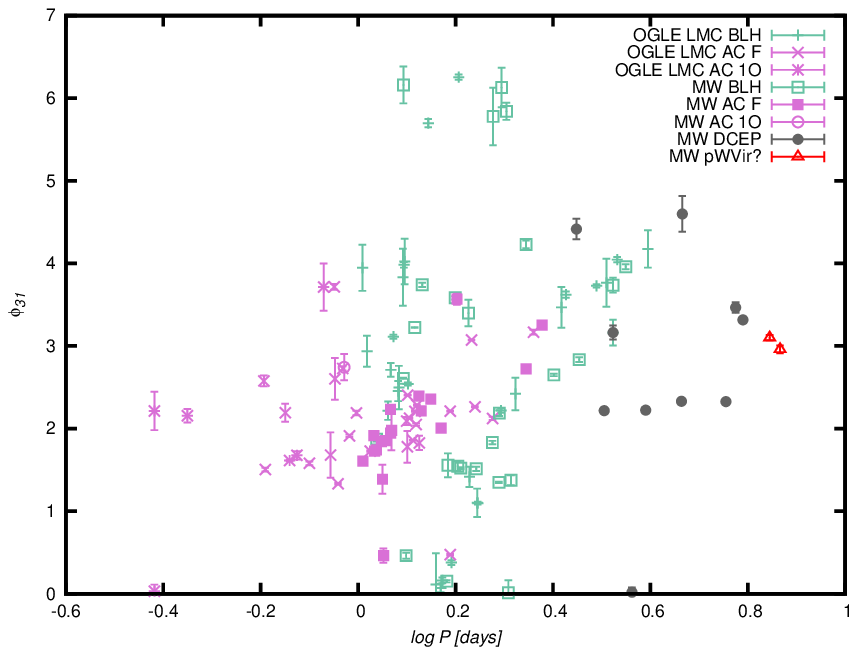}
\\
{(d)}
\end{minipage}

\figurecaption{1.}{The logarithm of the periods are plotter against the Fourier parameters $R_{21}$, $R_{31}$, $\phi_{21}$ and $\phi_{31}$ with their errors. The same color indicates the same classes for variables from the OGLE LMC and from the MW. The OGLE LMC BLHs are plotted in green crosses, the MW BLHs are green squares. The purple crosses represent the OGLE LMC AC F stars, the OGLE LMC AC 1O purple stars, the MW AC F are plotted as purple filled squares and MW AC 1O star is plotted as a purple circle. The MW DCEP are plotted as dark grey circles. Orange dotted triangles stand for possible pWVir stars.}

\begin{multicols}{2}


\centerline{\includegraphics[width=\columnwidth]{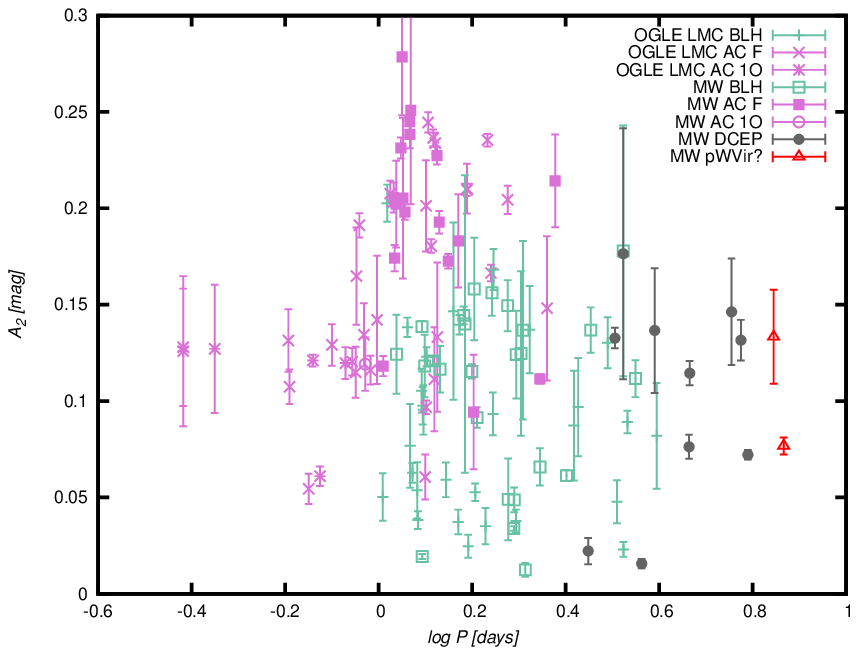}}
\figurecaption{2.}{Log $P$ vs. $A_{2}$ for the stars in the LMC. The description is the same as in Fig. 1.}   

\centerline{\includegraphics[width=\columnwidth]{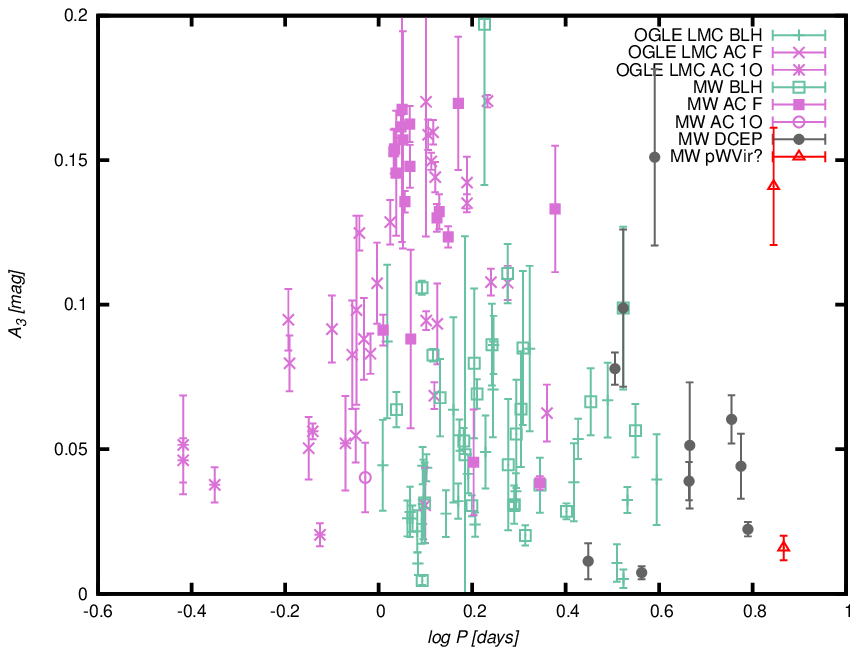}}
\figurecaption{3.}{Log $P$ vs. $A_{3}$ for the stars in the Field. The description is the same as in Fig. 1.}  


\end{multicols}
\begin{minipage}[b]{.24\linewidth}
\centering
\includegraphics[width=0.9\columnwidth, keepaspectratio]{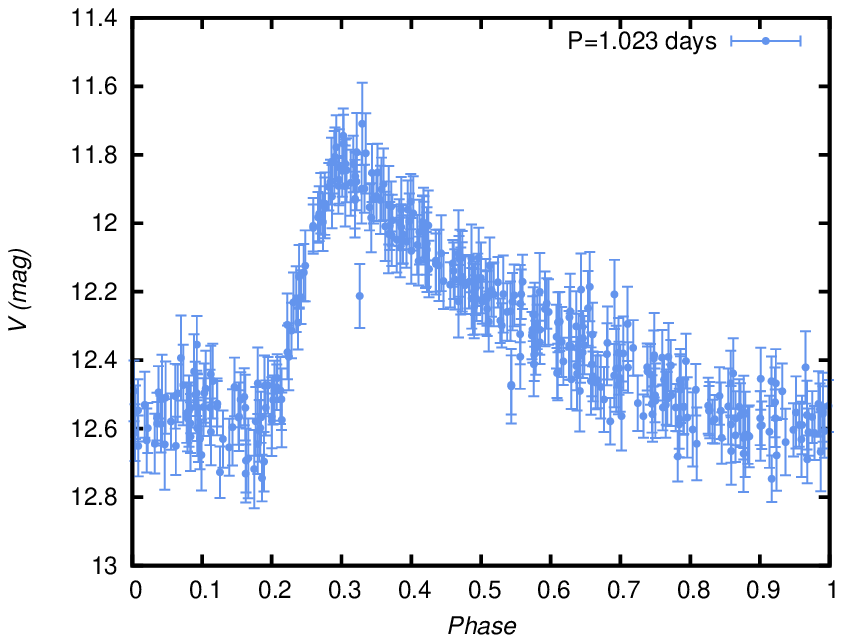}
\\
{(a) FY Aqr} 
\end{minipage}
\begin{minipage}[b]{.24\linewidth}
\centering
{\includegraphics[width=0.9\columnwidth, keepaspectratio]{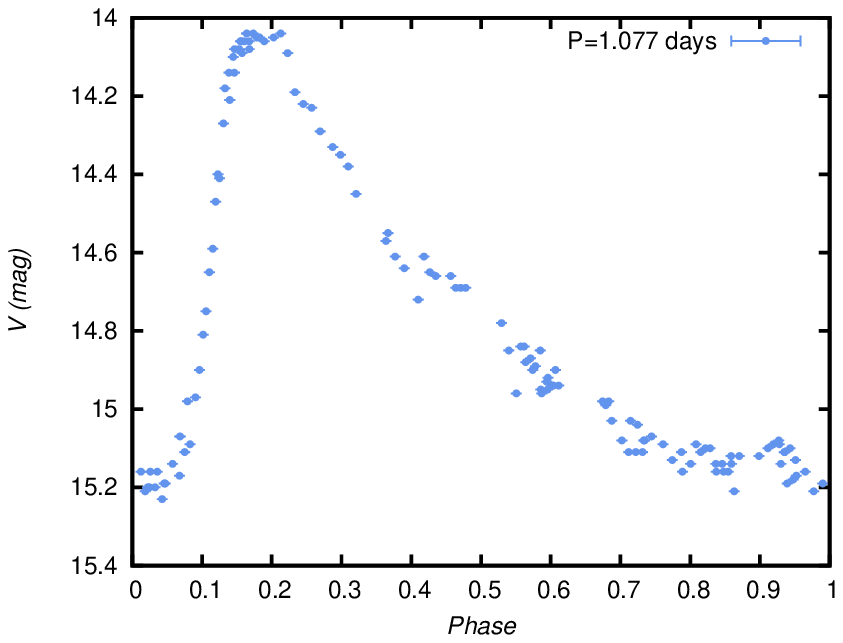}}
\\
{(b) V563 Cen} 
\end{minipage}
\begin{minipage}[b]{.24\linewidth}
\centering
{\includegraphics[width=0.9\columnwidth, keepaspectratio]{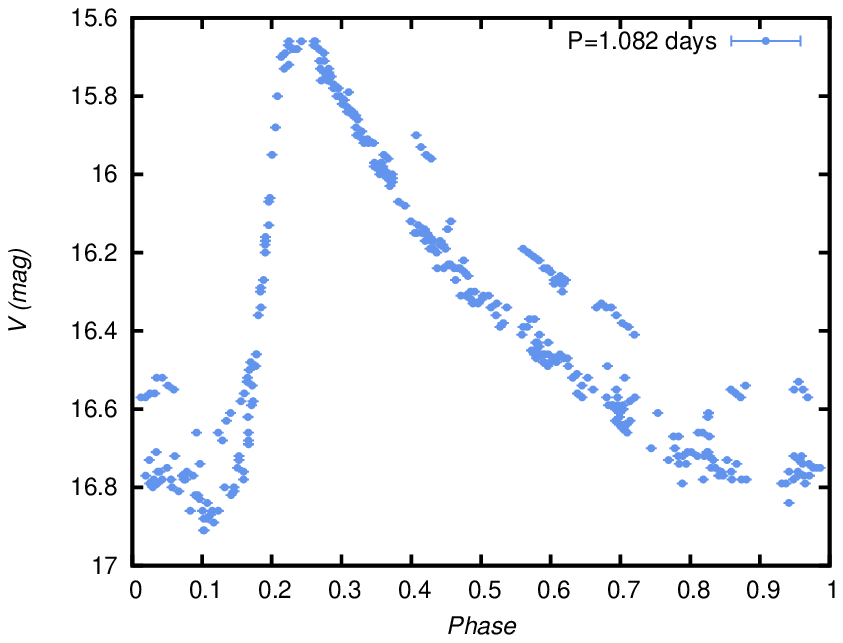}}
\\
{(c) FY Vir} 
\end{minipage}

\begin{minipage}[b]{.24\linewidth}
\centering
\includegraphics[width=0.9\columnwidth, keepaspectratio]{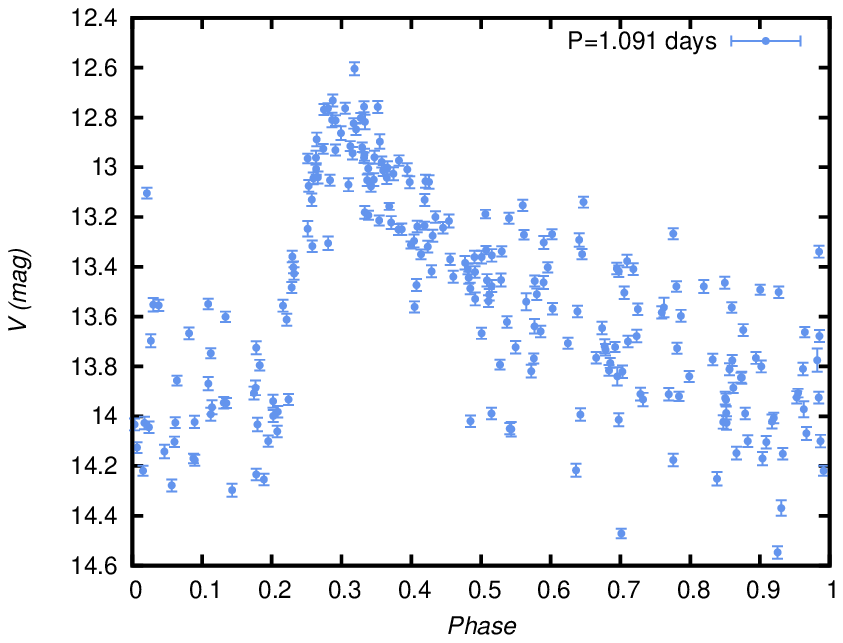}
\\
{(d) PP Tel} 
\end{minipage}
\begin{minipage}[b]{.24\linewidth}
\centering
{\includegraphics[width=0.9\columnwidth, keepaspectratio]{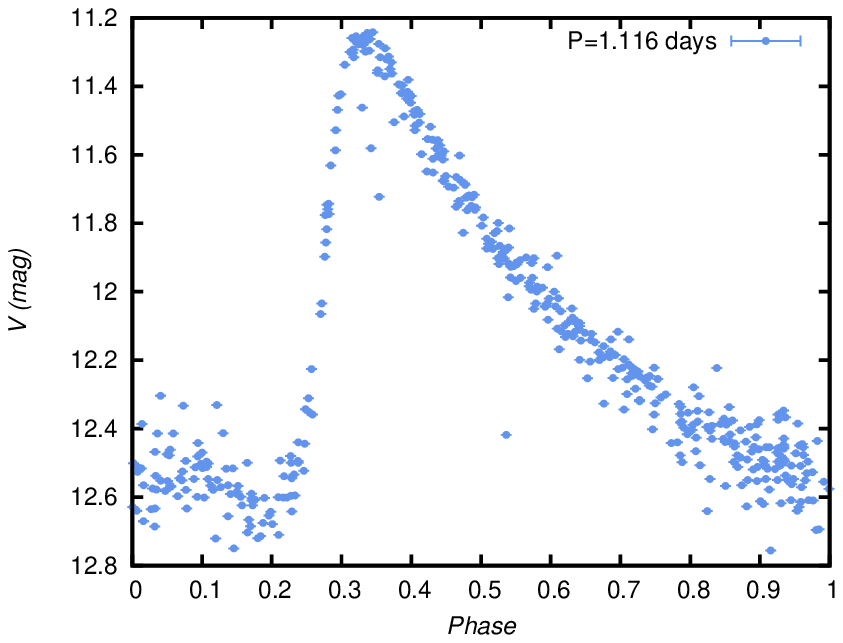}}
\\
{(e) V716 Oph} 
\end{minipage}
\begin{minipage}[b]{.24\linewidth}
\centering
{\includegraphics[width=0.9\columnwidth, keepaspectratio]{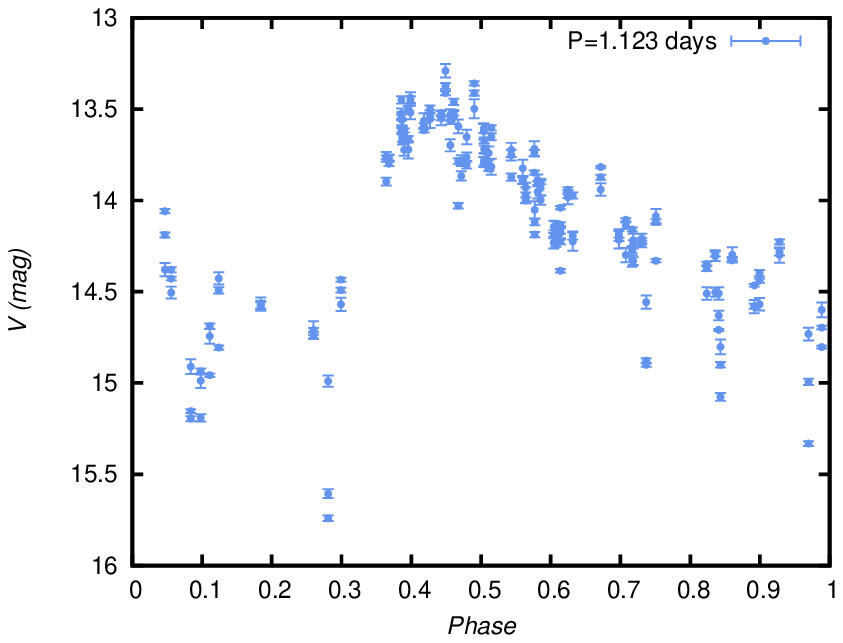}}
\\
{(f) DF Hyi} 
\end{minipage}

\begin{minipage}[b]{.24\linewidth}
\centering
\includegraphics[width=0.9\columnwidth, keepaspectratio]{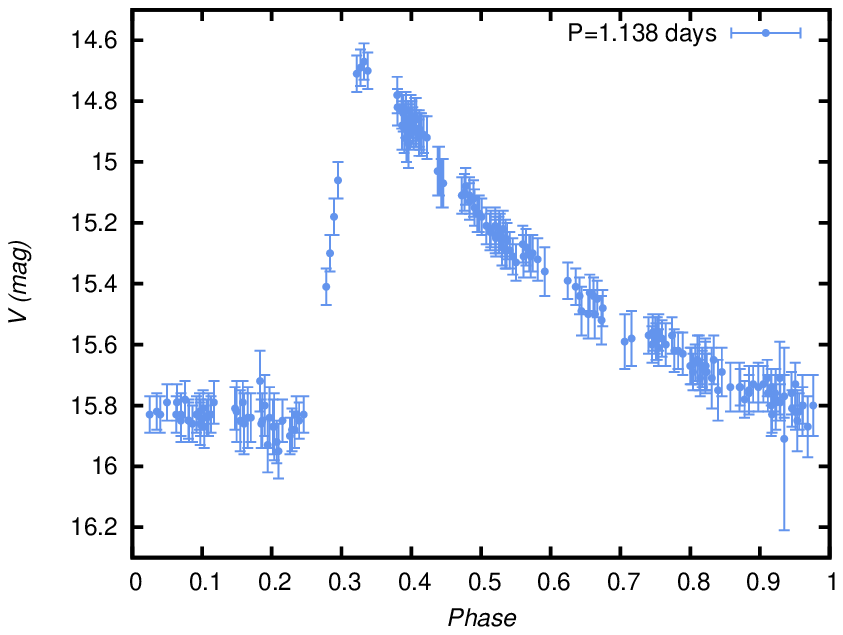}
\\
{(g) BH Cet} 
\end{minipage}
\begin{minipage}[b]{.24\linewidth}
\centering
{\includegraphics[width=0.9\columnwidth, keepaspectratio]{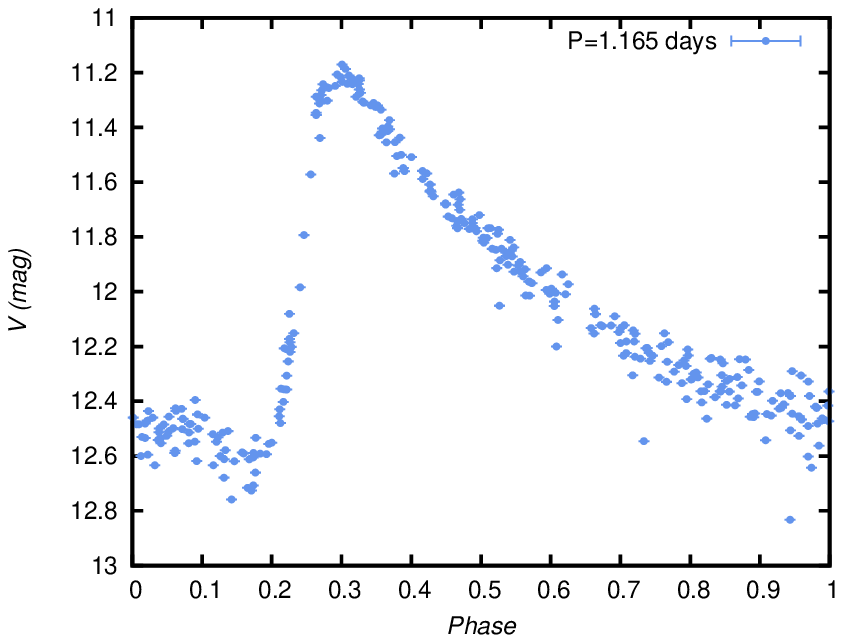}}
\\
{(h) BF Ser} 
\end{minipage}
\begin{minipage}[b]{.24\linewidth}
\centering
{\includegraphics[width=0.9\columnwidth, keepaspectratio]{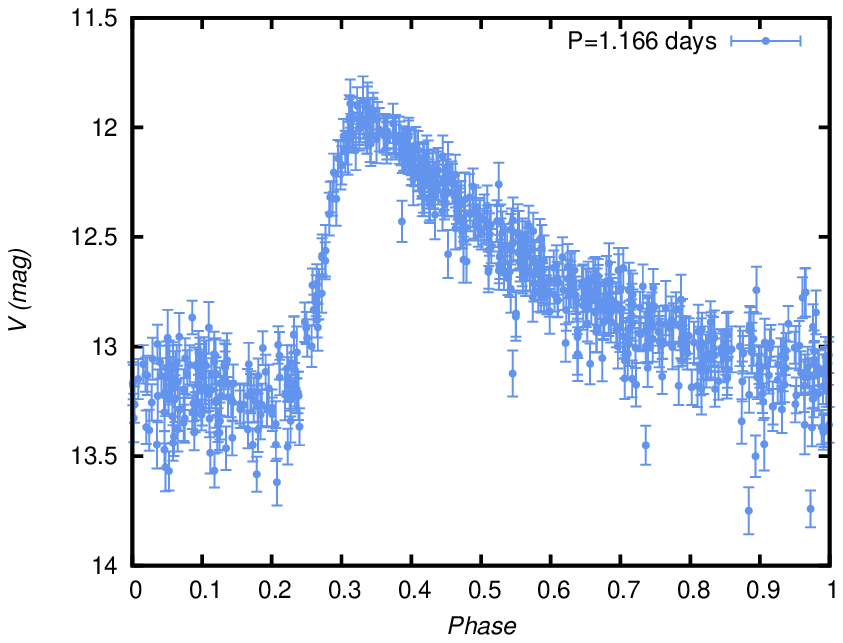}}
\\
{(i) BI Tel} 
\end{minipage}

\begin{minipage}[b]{.24\linewidth}
\centering
\includegraphics[width=0.9\columnwidth, keepaspectratio]{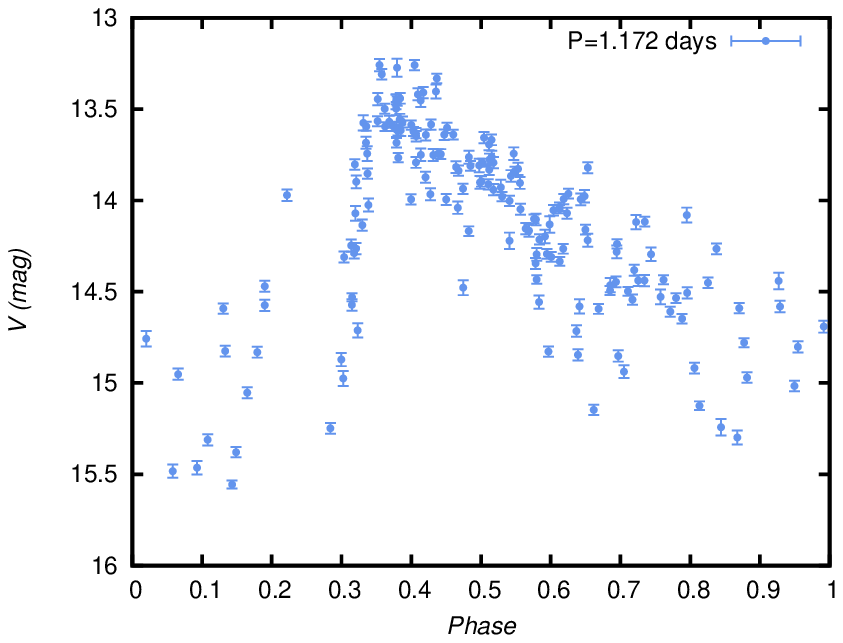}
\\
{(j) V2733 Oph} 
\end{minipage}
\begin{minipage}[b]{.24\linewidth}
\centering
{\includegraphics[width=0.9\columnwidth, keepaspectratio]{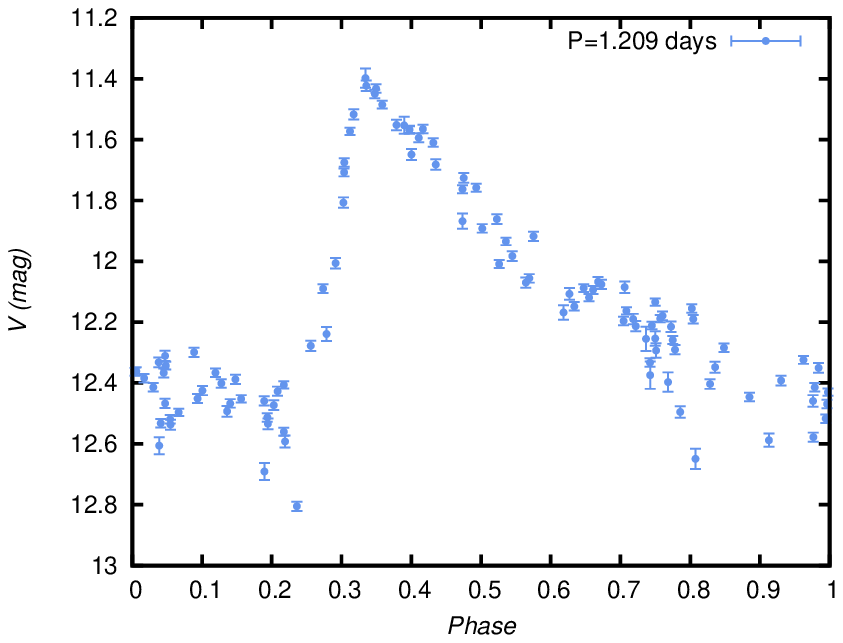}}
\\
{(k) CE Her} 
\end{minipage}
\begin{minipage}[b]{.24\linewidth}
\centering
{\includegraphics[width=0.9\columnwidth, keepaspectratio]{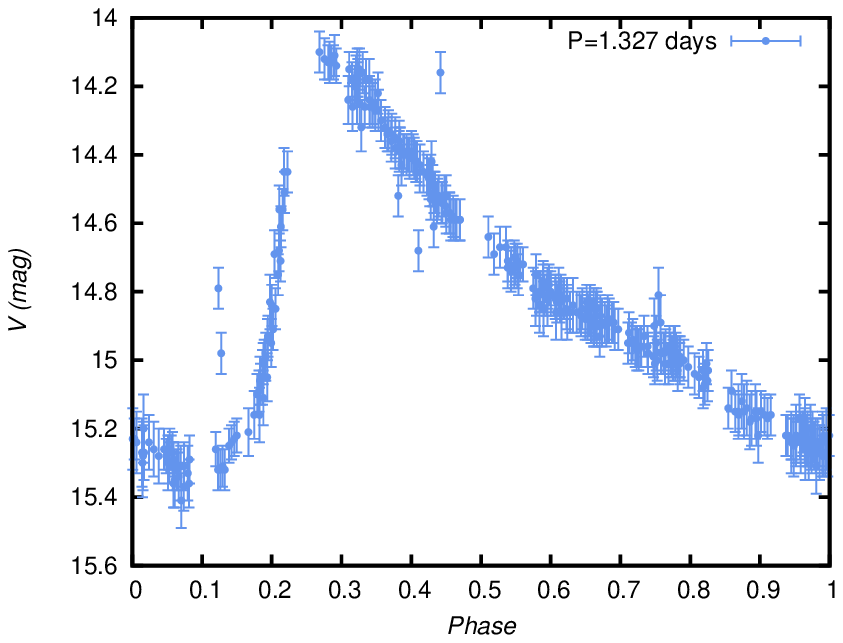}}
\\
{(l) VX Cap} 
\end{minipage}

\begin{minipage}[b]{.24\linewidth}
\centering
\includegraphics[width=0.9\columnwidth, keepaspectratio]{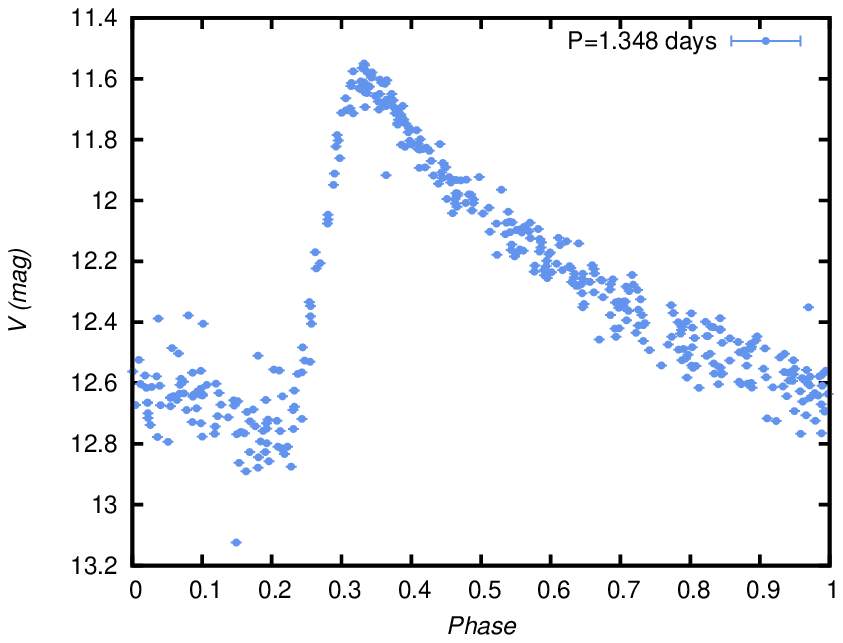}
\\
{(m) XX Vir} 
\end{minipage}
\begin{minipage}[b]{.24\linewidth}
\centering
{\includegraphics[width=0.9\columnwidth, keepaspectratio]{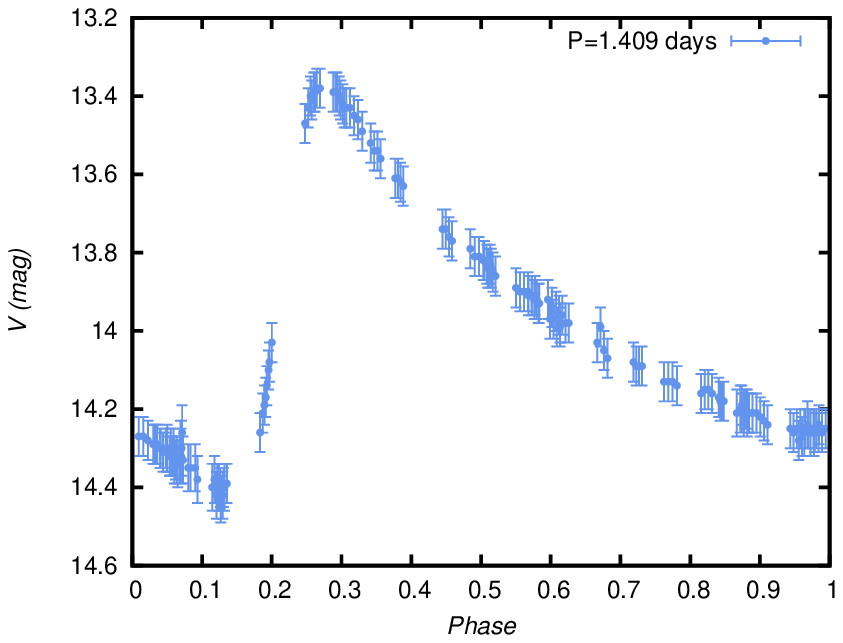}}
\\
{(n) V1149 Her} 
\end{minipage}
\begin{minipage}[b]{.24\linewidth}
\centering
{\includegraphics[width=0.9\columnwidth, keepaspectratio]{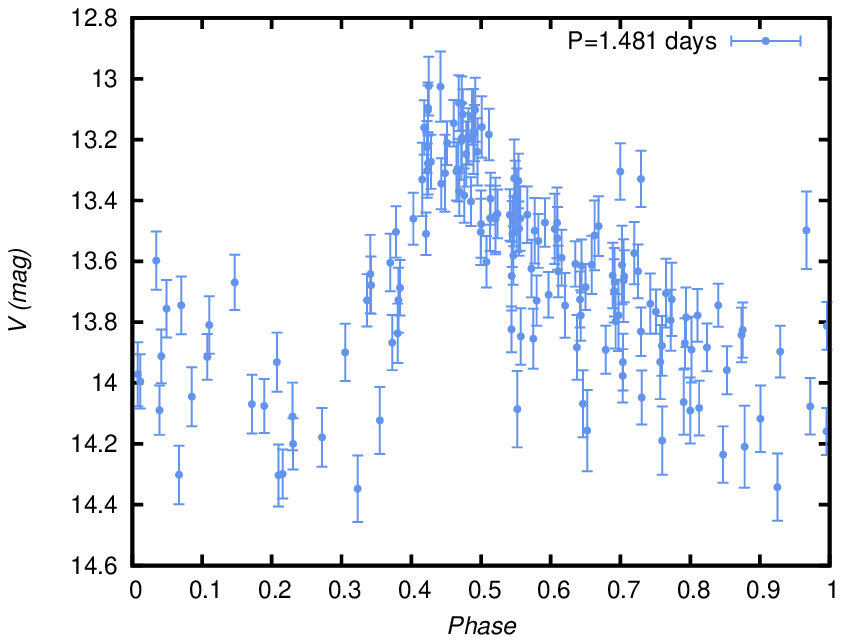}}
\\
{(o) MQ Aql} 
\end{minipage}

\begin{minipage}[b]{.24\linewidth}
\centering
\includegraphics[width=0.9\columnwidth, keepaspectratio]{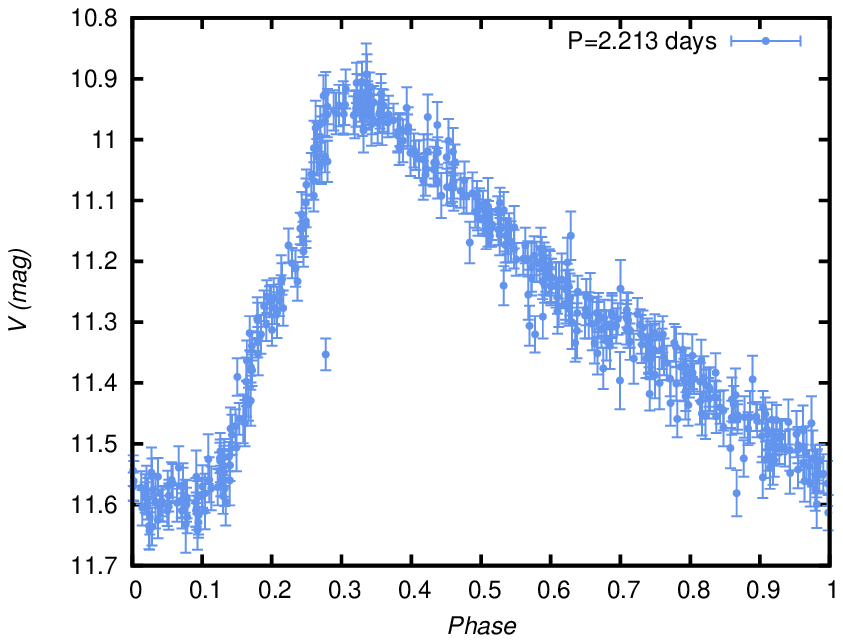}
\\
{(p) UY Eri} 
\end{minipage}
\begin{minipage}[b]{.24\linewidth}
\centering
{\includegraphics[width=0.9\columnwidth, keepaspectratio]{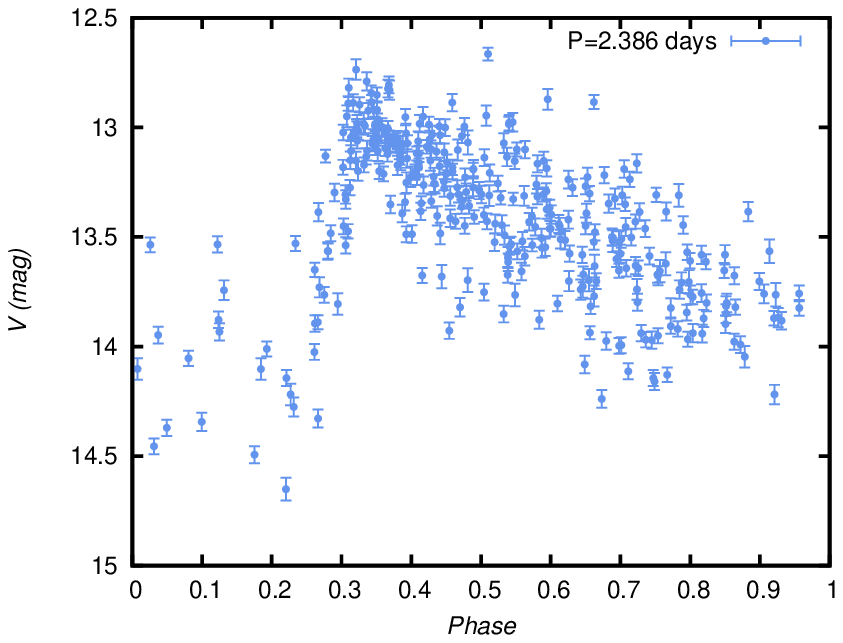}}
\\
{(q) UX Nor} 
\end{minipage}

\figurecaption{4.}{The phased light curves of ACs in the MW with photometric errors.}

\begin{multicols}{2}


\section{4. RESULTS}
\label{sec:results}

	Here we give the results of our analysis, separating each star into the their variable class, and giving additional details about them when necessary. It is important treat variable stars which are members of a globular or an open cluster separately from the individual stars in the MW, because we need to differentiate initial conditions at the birth of these stars, and the evolutionary processes that take place in these objects from those in the Galaxy. Contaminating the Galactic sample with these variables is going to influence our understanding of MW evolution itself, and lessen our knowledge about the clusters.


\subsection{4.1. The anomalous Cepheids}
\label{subsec:ACs}

	Soszy\'{n}ski et al. (2011a) and Soszy\'{n}ski et al. (2011b) show in their variable star distribution that the 1 days limit they have adopted for making a differentiation between the RRLs and T2Cs is well founded in the observations. The ACs bridge the span of both these variable types, as well as the DCEPs. The problem in the MW sample is that it is much more difficult to disentangle each of these variables on a $PL$ relation, since the reddening is changing significantly in different directions. This is why examination of the properties of the light curves is the best way to proceed. 

	The CSS (Drake et al. 2014a) used the Stetson variability index ($J_{WS}$) to identify the variability of the observed objects, and than applied a Lomb-Scargle periodogram analysis for the variables. The total number of ACs they have found is 64, and most of them are newly discovered variables, but six overlap with our sample: FY Vir, V716 Oph, BF Ser, VX Cap, XX Vir, V1149 Her. We confirmed these stars. In Drake et al. (2017) CSS published a list of 156 ACs in which two more stars were overlapping: V563 Cen and BI Tel. From our Fourier parameter plots we have found additional fundamental mode ACs: FY Aqr, PP Tel, DF Hyi, BQ CrA, BH Cet, V2733 Oph, CE Her, MQ Aql, V745 Oph, UY Eri, UX Nor, and one possible first overtone AC: V742 Cyg, see Figs. 4 and 5.

	We present the list of ACs with their positions (RA (h:m:s), DEC (d:m:s), eq=J2000) taken from Simbad\footnote{http://simbad.u-strasbg.fr}, the derived median V magnitude with additional data from the literature $T_{\rm eff}$, $logg$, [Fe/H], and the classification from the CSS in Table 1.

	The detailed Fourier parameters can be found in appendix Table A.1. 

	The phased light curves of AC stars are presented in Fig. 4.


	There are individual objects that we needed to expand on, because their interpretation needs further explaining.
	
	{\bf FY Vir} 
	
	The phased light curve of FY Vir ($P$=1.082 days) seen in Fig. 4c shows a well defined modulation. 
	

	{\bf V742 Cyg}
	
	Zejda et al. (2012) states that V742 Cyg ($P$=0.936 days) is a member of the open cluster Dolidze 37. 
	The data for our analysis was published by Schmidt and Reiswig (1993). In the VizieR catalog the data for V742 Cyg was wrongly published under the name "V741 Cyg", which is a known eclipsing binary of Algol type with a period of $P$=4.649850 days. The correctly phased light curve can be see in Fig. 5. Vasil'Yanovskaya (1978), Schmidt (1990), and Schmidt (1991) all discussed that V742 Cyg shows a significant period variation (with a period around $P$=0.93979 days) that is different from the Blazhko effect in RRL stars. 	
	Based on the Fourier parameters it is an AC, possibly a 1O AC, making this star the first possible AC to be discovered in an open cluster. 
	
\centerline{\includegraphics[width=\columnwidth]{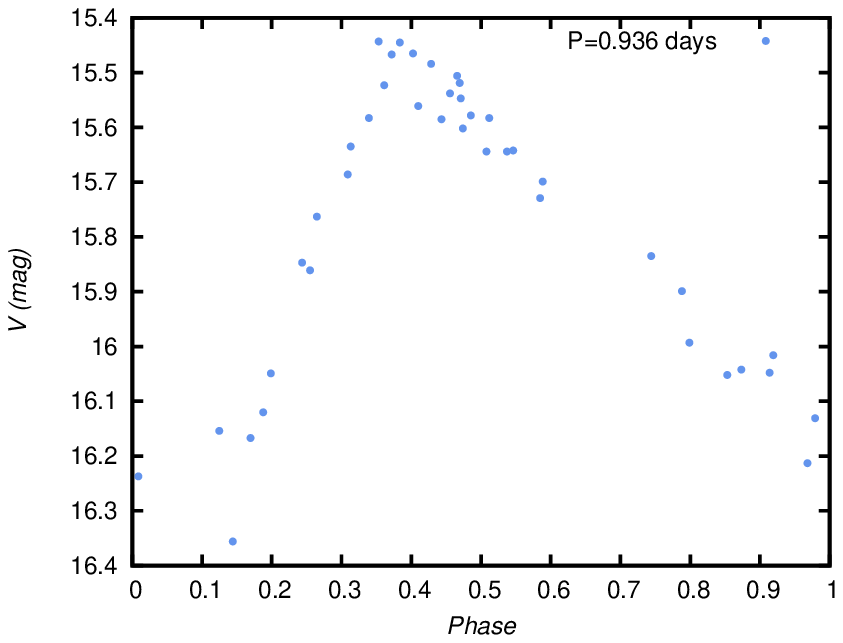}}
\figurecaption{5.}{V742 Cyg phased light curve with a $P$=0.936 days.}


	{\bf V716 Oph}
	
	V716 Oph ($P$=1.116 days) is a member of the globular cluster $\omega$ Centauri (NGC 5139), Dinescu (2002). The estimated age of $\omega$ Cen is 11.52$\times$10$^9$ yrs. Resulting from our Fourier parameters V716 Oph is an AC, confirming the results of Drake et al. (2014a).
	

	{\bf BI Tel}

	The classification of this object has been shifting during the years. Layden (1994) lists BI Tel ($P$=1.166 days) as an "RRab" type RRL, and get the [Fe/H]=-1.96 dex for the metallicity. Than with the ASAS-3 observations, the classification changed to "DCEP-FU" (Pojmanski et al. 2005), to be reclassified, again, to an "RR Lyrae FM" by Richards et al. (2012). Our Fourier parameters show it to be an AC.


	{\bf BQ CrA and V745 Oph}
	
	BQ CrA ($P$=1.128 days) and V745 Oph ($P$=1.596 days) have light curves that show a lot of scatter (see Fig. 6), so their new classification may be wrong, and needs further confirmation.

\end{multicols}
\begin{minipage}[b]{.4\linewidth}
\centering
\includegraphics[width=0.9\columnwidth, keepaspectratio]{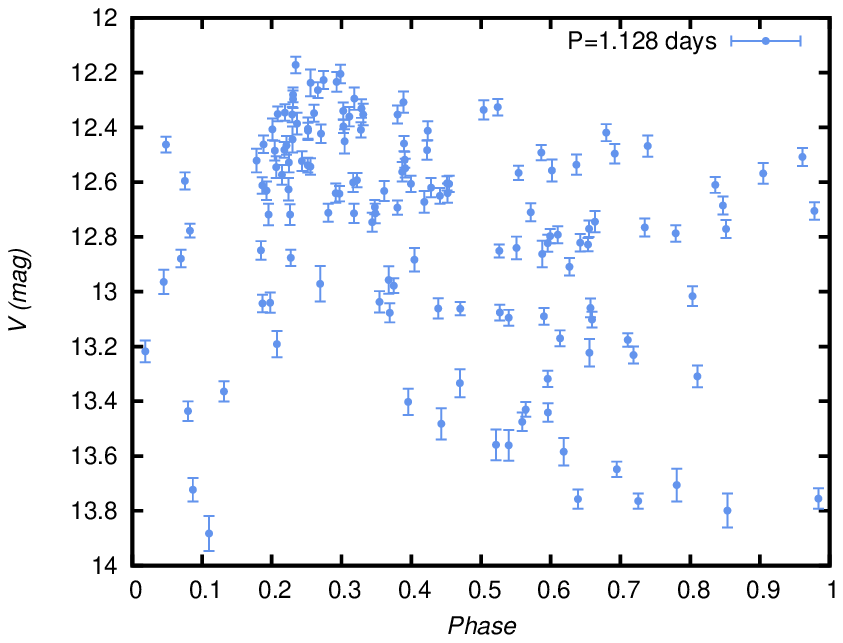}
\\
{(a) BQ CrA} 
\end{minipage}
\begin{minipage}[b]{.4\linewidth}
\centering
{\includegraphics[width=0.9\columnwidth, keepaspectratio]{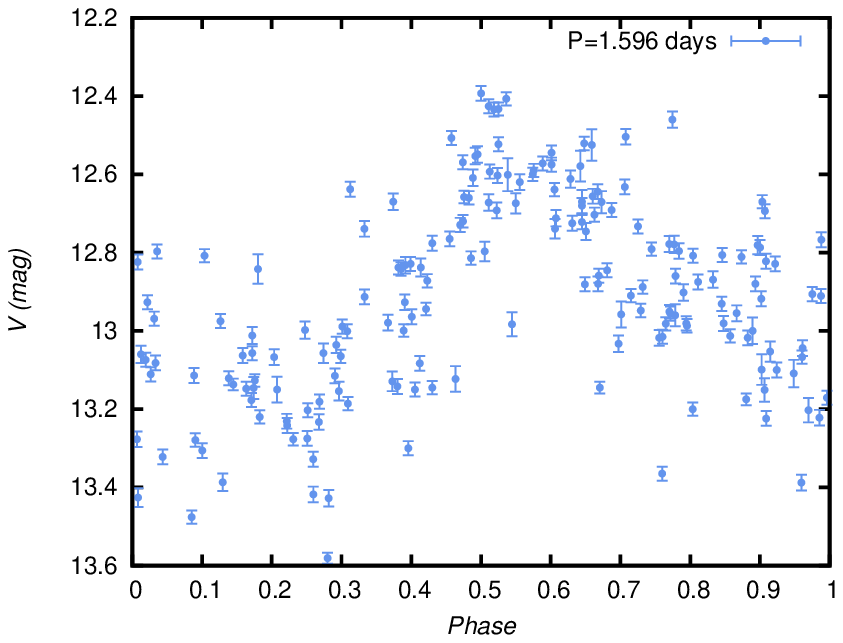}}
\\
{(b) V745 Oph} 
\end{minipage}

\figurecaption{6.}{The phased light curves of BQ CrA (left) and V745 Oph (right) showing significant scatter from ASAS-3 data.}

\begin{multicols}{2}


\end{multicols}

\noindent
\parbox{\textwidth}{
{\bf Table 1.} The identified ACs among the BLHs in the MW.
\vskip.25cm \centerline{
\begin{tabular}{l l l p{1.1cm} l p{0.5cm} p{0.7cm} p{1.cm} p{2.3cm} l}
\hline\
Name & RA (h:m:s) & DEC (d:m:s) & $<V>$ [mag] &  $P$ [days] & $T_{\rm eff}$ [K] & $logg$ [cgs] & [Fe/H] [dex] & References & CSS \\
\hline
V742 Cyg & 20:02:32.77 & +37:46:34.8 & 15.886 & 0.936 & & & & & \\
FY Aqr & 22:16:34.99 & -03:48:55.41 & 12.359 & 1.023 & & & & & \\
V563 Cen & 14:49:50.89 & -38:53:08.4 & 14.798 & 1.077 & & & & & AC\\
FY Vir & 12:14:13.52 & +06:01:17.1 & 16.419 & 1.082 & & & & & AC\\
PP Tel & 20:16:56.51 & -51:15:11.4 & 13.622 & 1.091 & & & & & \\
V716 Oph & 16:30:49.47 & -05:30:19.5 & 12.169 & 1.116 & 6550 & 2.500 & -1.870 & Soubiran et al. (2016) & AC\\
DF Hyi & 01:40:49.20 & -67:29:41.9 & 14.336 & 1.123 & & & & & \\
BQ CrA & 18:37:01.38 & -38:04:34.5 & 13.132 & 1.128 & & & & & \\
BH Cet & 00:50:02.80 & -17:36:26.9 & 15.490 & 1.138 & & & & & \\
BF Ser & 15:16:28.50 & +16:26:39.69 & 12.093 & 1.165 & & & & & AC\\
BI Tel & 18:18:17.56 & -53:22:19.8 & 12.814 & 1.166 & - & - & -1.96 & Layden (1994) & AC\\
V2733 Oph & 17:31:21.880 & -17:43:39.51 & 14.469 & 1.172 & & & & & \\
CE Her & 17:41:56.55 & +15:04:30.2 & 12.170 & 1.209 & & & & & \\
VX Cap & 21:06:22.51 & -18:49:39.8 & 14.859 & 1.327 & & & & & AC\\
XX Vir & 14:16:48.59 & -06:17:15.06 & 12.326 & 1.348 & & & & & AC\\
V1149 Her & 16:03:43.36 & +50:13:33.43 & 13.995 & 1.409 & - & - & -2.32 & Allende Prieto et al. (2000) & AC\\
MQ Aql & 19:40:55.69 & +12:37:10.1 & 13.812 & 1.481 & & & & & \\
V745 Oph & 17:20:02.93 & +03:48:56.0 & 13.942 & 1.596 & & & & & \\
UY Eri & 03:13:39.13 & -10:26:32.40  & 11.294 & 2.213 & 6800 & 1.800 & -1.430 & Soubiran et al. (2016) & \\
 & & & & & 6389 6280 9978 & 3.25 2.58 4.15 & 0.01 0.10 0.30 & Schmidt et al. (2011) & \\
 & & & & & 6000 & 1.5 & -1.8 & Maas et al. (2007) & \\
UX Nor & 16:27:44.70 & -56:47:08.1 & 13.644 & 2.386 & & & & & \\	
\hline
\end{tabular}
}  
}  
\vskip.5cm

\begin{multicols}{2}


	{\bf XX Vir}
	
	Dr Dana Casetti in private communication confirmed that XX Vir ($P$=1.348 days) is not a member $\omega$ Centaur globular cluster. We agree that XX Vir is an AC, as identified in Drake et al. (2014a).


	{\bf UY Eri}
	
	The results of the three different articles (Maas et al. 2007, Schmidt et al. 2011, Soubiran et al. 2016) that measured metallicity for UY Eri ($P$=2.213 days) are summarized in Table 1. The big differences in the [Fe/H] values make conclusions about this star open for further discussion. While on most of the Fourier parameter plots UY Eri is among AC, in amplitude figures (see Figs. 2 and 3) it seems to be among the BLHs. Further measurements are needed to resolve the question about the nature of this star.


\subsection{4.2. The short period Type II Cepheids - BL Herculis subtype}
\label{subsec:BLH}

	Summarizing the results from the Fourier parameter plots we list the BLHs with additional information: positions (RA (h:m:s), DEC (d:m:s), eq=J2000 from Simbad\footnote{http://simbad.u-strasbg.fr}, the derived median V magnitude, and $T_{\rm eff}$, log g, [Fe/H] from the literature) in Table 2. All, but two, stars had their [Fe/H] determined from direct spectroscopic measurements. The exceptions are V527 Sgr ($M$=1.025 M$_{\odot}$, $R$=1.366 R$_{\odot}$) and V1287 Sco ($M$=0.963 M$_{\odot}$, $R$=2.933 R$_{\odot}$), for which the parameters were derived with a combination of photometric observations and a MW stellar population synthesis program (Sharma et al. 2011) for the EPIC K2 stars (Huber et al. 2016). 

	There are some BLH stars in our sample that need more detailed discussion, which we give in the following paragraphs.

	{\bf BL Her}
	
	The literature dealt with BL Her ($P$=1.307 days) in quite some detail. The behaviour of the H$_{\alpha}$ line was analysed by Gillet et al. (1994), while Fokin and Gillet (1994) compared the measurements with a non-linear model, to find a good agreement and confirmation of the presence of the 2:1 resonance predicted by the pulsation model of T2Cs. The emission in the H$_{\alpha}$ line was reported in Schmidt et al. (2003) confirming previous findings of Gillet et al. (1994) and Vinko et al. (1998) (and references within). The numerous measurements of the $T_{\rm eff}$, log g and [Fe/H] from spectra are listed in Table 2. Furthermore, a Baade-Wesselink analysis carried out by Balog et al. (1997) on BL Her gave the following results: $R$=9.4 $\pm$ 2 R$_{\odot}$. Using the surface-brightness method for deriving radii of stars Arellano Ferro and Rosenzweig (2000) derived $R$=15.4 $\pm$ 1.5 R$_{\odot}$, while Groenewegen and Jurkovic (2017b) used the period-radius relation to get $R$=8.066 R$_{\odot}$, putting it outside the region of the BLHs.  
	On the Fourier parameters figures the position of BL Her is not unambiguous, either. The true nature of BL Her is still open for discussion, despite the fact that this star named the whole subgroup.


	{\bf KZ Cen}
	
	The Fourier parameters and light curve shape puts KZ Cen ($P$=1.520 days) among BLHs, but the metallicity given in Table 2 is 0.67, which is too high for this subtype, so this makes KZ Cen great candidate for further spectroscopic measurements. 
	

\end{multicols}

\noindent
\parbox{\columnwidth}{
{\bf Table 2.} The list of BLHs.
\vskip.25cm \centerline{
\resizebox{\linewidth}{!}{
\begin{tabular}{l l l l l p{1.3cm} p{1.0cm} p{1.2cm} l p{3.5cm}}
\hline\
Name & RA (h:m:s) & DEC (d:m:s) & $<V>$ [mag] &  P [days] & $T_{\rm eff}$ [K] & log g [cgs] & [Fe/H] [dex] & References & Notes \\
\hline
BX Del & 20:21:18.97 & +18:26:16.28 & 12.412 & 1.092 & 6250 & 1.0 & -0.2 & Maas et al. (2007) & \\
BV Cha & 13:02:21.18 & -79:45:26.4 & 12.284 & 1.238 & & & & & \\
VY Pyx & 08:54:29.63 & -23:31:18.57 & 7.245 & 1.240 & 5750 & 1.5 & -0.4 & Maas et al. (2007) & \\
V527 Sgr & 19:16:10.99 & -20:55:55.8 & 14.918 & 1.255 & 5816 & 4.159 & -0.042 & Huber et al. (2016) using Sharma et al. (2011) & \\
BL Her & 18:01:09.22 & +19:14:56.70 & 10.219 & 1.307 & 6121 & - & - & Mu\~{n}oz Bermejo et al. (2013) & \\
& & & & & 6256 & - & - & McDonald et al. (2012) & \\
& & & & & 6500 & 2.0 & -0.1 & Maas et al. (2007) & \\
& & & & & 6497 6464 & - & -0.17 & Ammons et al. (2006) & \\
& & & & & 6350 & 2.5 & 0.0 & Caldwell and Butler (1978) & \\
V5614 Sgr & 17:55:43.81 & -29:44:50.2 & 16.656 & 1.354 & & & & & \\
HQ CrA & 18:11:53.74 & -37:39:15.0 & 14.741 & 1.415 & & & & & Madore and Fernie (1980): blue companion?\\
KZ Cen & 12:01:55.19 & -46:16:41.48 & 12.293 & 1.520 & 6261 6021 & - & 0.67 & Ammons et al. (2006) & \\
V2022 Sgr & 18:40:39.0 & -25:22:50 & 13.516 & 1.529 & & & & & \\
SW Tau & 04:24:32.97 & +04:07:24.08 & 9.721 & 1.584 & 7036 7482 6322 & 3.10 3.54 1.92 & 0.20 0.22 0.12 & Schmidt et al. (2011) & \\
 & & & & & 6250 & 2.0 & 0.2 & Maas et al. (2007) & \\
NW Lyr & 19:15:56.34 & +34:27:08.08  & 12.520 & 1.601 & & & & & \\
VZ Aql & 19:05:02.96 & -06:50:58.7 & 14.002 & 1.685 & & & & & \\
V1437 Sgr & 18 03 33.63 & -30 01 14.9 & 15.603 & 1.748 & & & & & \\ 
V714 Cyg & 19:41:48.62 & +37:59:33.8 & 14.148 & 1.888 & & & & & \\
RT TrA & 16:34:30.89 & -63:08:00.81 & 9.841 & 1.946 & 5996 5868 & - & 0.48 & Ammons et al. (2006) & C-rich\\
& & & & & 6200 6040 6360 & 2.0 2.5 2.3 & 0.34 0.54 0.43 & Wallerstein et al. (2000) & \\
V439 Oph & 17:43:33.27 & +03:35:36.08 & 12.172 & 1.893 & 5547.61 & & -0.493 & Luo et al. (2016) & \\
GK Cen & 13:46:20.94 & -49:35:50.1 & 12.929 & 1.950 & & & & & \\
V477 Oph & 17:59:08.16 & +05:38:26.2 & 13.946 & 2.016 & & & & & \\
V1287 Sco & 16:36:52.85 & -28:05:34.2 & 13.480 & 2.036 & 5428 & 3.542 & -0.343 & Huber et al. (2016) using Sharma et al. (2011) & period increase\\
AT Tel & 18:50:02.64 & -51:38:04.6 & 14.216 & 1.970 & & & & & \\
V553 Cen & 14:46:33.64 & -32:10:15.25 & 8.458 & 2.061 & 5600 & 3.100 & -0.500 & Soubiran et al. (2016) & C-rich\\ 
& & & & & 5635 5654 & - & 0.16 & Ammons et al. (2006) & \\
& & & & & - & - & 0.04 & Wallerstein and Gonzalez (1996) & \\
& & & & & 5600 & 3.1 & -0.5 & Wallerstein et al. (1979) & \\	
V5608 Sgr & 17:54:09.11 & -29:39:59.0 & 16.374 & 2.212 & & & & & \\
V617 Ara & 17:10:08.31 & -60:39:43.4 & 11.792 & 2.522 & & & & & DCEP in \textit{Simbad} \\
V465 Oph & 17:52:07.5 & -01:05:07 & 13.312 & 2.844 & & & & & \\
BE CrA & 18:54:54.03 & -40:23:15.3 & 13.957 & 3.337 & & & & & \\
V5609 Sgr & 17:54:55.57 & -29:57:31.0 & 16.919 & 3.542 & & & & & \\
\hline
\end{tabular}
}  
}  
}
\vskip.5cm

\begin{multicols}{2}


	{\bf V2022 Sgr} 
	
	V2022 Sgr ($P$=1.529 days) was observed by Diethelm (1983), who found a period of $P$=1.533 days, while Kwee and Diethelm (1984) observed it in UBV-bands determining a $P$=1.533171 days. Provencal (1986) has observed that the period has increased up to $P$=1.5530160 days. 
	
	In Jurkovic (2015) V2022 Sgr was misclassified. The confusion came from the reference on the "The ASAS Catalogue of Variable Stars" which directs to this object: 184041-2523.7. This object is indeed a variable star with a period of $P$=290.71001 $\pm$ 0.52327 days. The right ASAS ID for V2022 Sgr is 184039-2523.0, and it appears correctly in the "ASAS All Star Catalogue". 
	
	We can confirm that V2022 Sgr is a BLH, with a variable period, using data from Kwee and Diethelm (1984).

	
	{\bf V1437 Sgr}

	V1437 Sgr ($P$=1.748 days) is a probable member of NGC~6522 (V8) globular cluster according to the newest catalogue of variables in clusters by Clement (2017) (for details see Clement et al. (2001)). This is also confirmed by Samus et al. (2009b). Udalski et al. (1994) in the VizieR catalogue (Udalski 1996) the OGLE star identifier for this object is BWC V1, and gives a type for the variability type "ACEP", and it is also remarked that this star was identified by Blanco (1984) as a "CW" and marked V58. The light curve is (and I, V measurements) can be found in Soszy\'{n}ski et al. (2011b), where it is, once again, classified as a T2C-BLH. We confirm this latest result with the Fourier parameters.
	
	
	{\bf RT TrA}

	RT TrA ($P$=1.946 days) is a carbon rich star (Lloyd Evans 1983, Wallerstein et al. 2000, Wallerstein 2002), as is V553 Cen, with a light curve (Diethelm 1983) that puts its among BLH (see Fig. 7a), but with a bump on the ascending branch. It looks similar to a BLH star with a similar period, but this feature makes it different. The Fourier parameters (as with the V553 Cen) the $R_{21}$, $R_{31}$ is very close to 0. The evolutionary models do not give an answer for the existence of such C-rich stars, as it is detailed in Wallerstein (2002).
	
\end{multicols}
\begin{minipage}[b]{.4\linewidth}
\centering
\includegraphics[width=0.9\columnwidth, keepaspectratio]{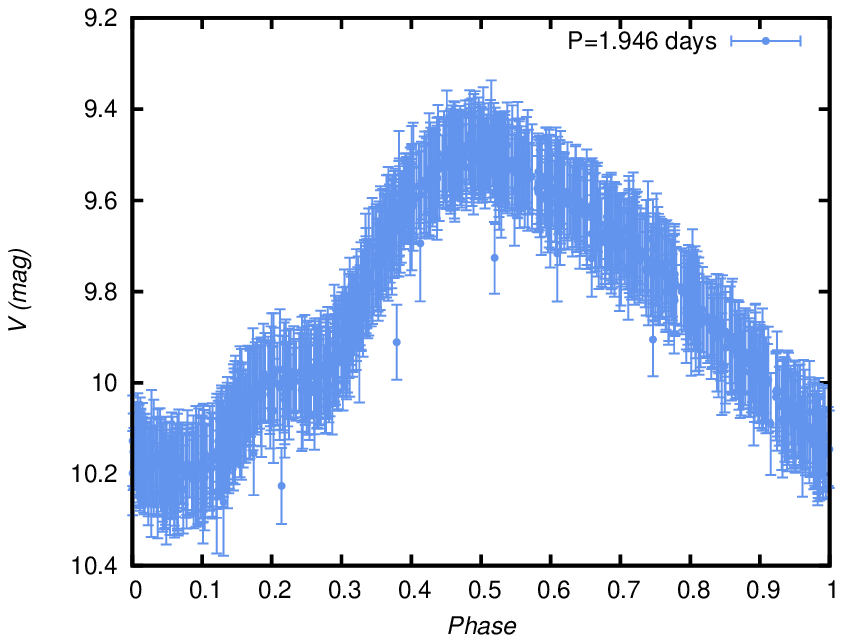}
\\
{(a) Phased RT TrA light curve from ASAS-3 data ($P$=1.946 days).} 
\end{minipage}
\begin{minipage}[b]{.4\linewidth}
\centering
{\includegraphics[width=0.9\columnwidth, keepaspectratio]{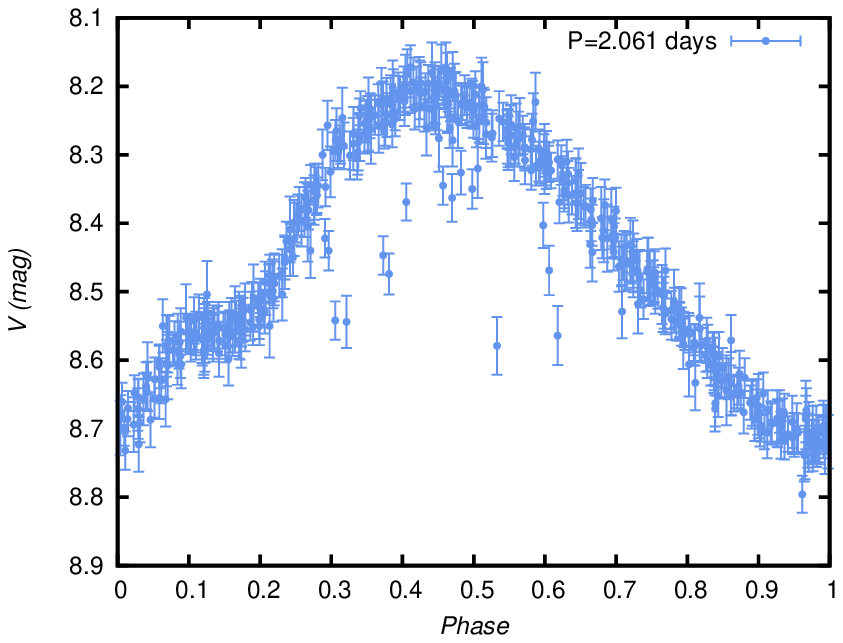}}
\\
{(b) Phased light curve of V553 Cen from ASAS-3 data ($P$=2.061 days).} 
\end{minipage}

\figurecaption{7.}{The phased light curves of the C-rich BLH stars.}

\begin{multicols}{2}	
	

	{\bf V477 Oph}
	
	V477 Oph ($P$=2.016 days) is a member of the open cluster Collinder 359 according to (Zejda et al. 2012). If it is truly a cluster member this star can not be a BLH star, because the age of the cluster is much younger ($\approx$ 30-60 $\times$ 10$^6$ years, see Bobylev (2008)) than any BLH star should be.
	
	
	{\bf V553 Cen}
	
	Wallerstein and Gonzales (1996) has confirmed from a detailed spectroscopic study that V553 Cen ($P$=2.061 days) is indeed a C- and N-rich short period T2C, and in Wallerstein (1979) presented the measured $T_{\rm eff}$=5600 K, $logg$=3.1, and [Fe/H]=-0.50 (see Table 2). It is in the same time a O-poor star, and has a moderate Na content. The idea of this BLHs star being in a binary system was tested by Wallerstein and Gonzales (1996), but no evidence has been found that would support that. The only strange thing that is different from the other BLHs is that the $R_{21}$ and $R_{31}$ parameters are almost zero, a feature seen in the other C-rich star in the sample, RT TrA. Fig. 7 shows both RT TrA and V553 Cen showing very similar light curve shapes.


\subsection{4.3. The (p) W Virginis subtype Type II Cepheids}
\label{subsec:pWVir}

	The peculiar W Virginis (pWVir) stars are a special class among the W Virginis (WVir) with periods between 4 and 20 days. Many of the pWVir in the OGLE-III LMC sample are binary systems. They might overlap with the DCEPs on the $PL$ relations, because they are usually brighter than regular WVir stars. The phased light curves of UY CrA and IT Cep are shown in Fig. 8.
	
	{\bf UY CrA}
	
	The INTEGRAL data of UY CrA ($P$=6.995 days) is quite noisy, but we could derive the Fourier parameters, which puts this star among DCEP/pWVir stars, although the shape of the light curve would support the suggestion that it would be more likely that it a DCEP. 
	

	{\bf IT Cep}

	From the Fourier parameters IT Cep ($P$=7.349 days) can be classified as pWVir. Ammons et  al. (2006) lists the $T_{\rm eff}$=6622 and 6497 K, and a [Fe/H]=0.16 [dex]. 
	
\end{multicols}
\begin{minipage}[b]{.4\linewidth}
\centering
\includegraphics[width=0.9\columnwidth, keepaspectratio]{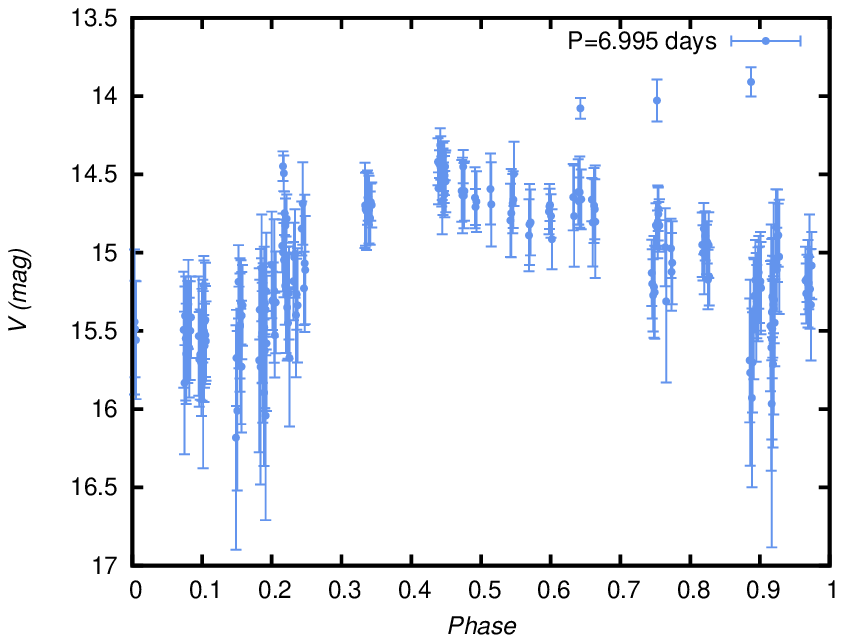}
\\
{(a) UY CrA phased with $P$=6.995 days.} 
\end{minipage}
\begin{minipage}[b]{.4\linewidth}
\centering
{\includegraphics[width=0.9\columnwidth, keepaspectratio]{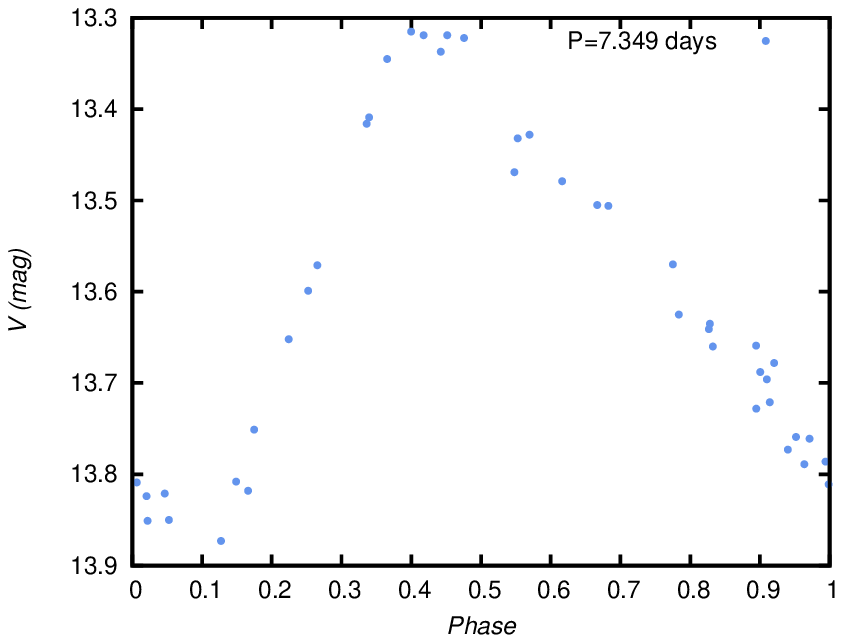}}
\\
{(b) IT Cep phased with $P$=7.349 days.} 
\end{minipage}

\figurecaption{8.}{The phased light curves of the suspected pWVir/DCEP stars - UY CrA on the left and IT Cep on the right.}

\begin{multicols}{2}	


\subsection{4.4. The classical Cepheids}
\label{subsec:DCEP}

	Classical Cepheids (DCEPs) are intermediate mass (4-20 M$_\odot$) stars pulsating in F, 1O and second overtone (2O) modes with periods from 1 to 100 days (Catelan and Smith 2015). The DCEPs that were found during our investigation (combining the Fourier parameters, and the light curve shapes) were summarized in Table 3 (RA (h:m:s), DEC (d:m:s), eq=J2000 from Simbad\footnote{http://simbad.u-strasbg.fr}, and derived media V magnitude). Where available we give the known $T_{\rm eff}$, $logg$ and [Fe/H] from spectroscopic measurements, as well as other physical parameters found in the literature when describing individual stars. 

	In the cases of V5626 Sgr (OGLE BUL-SC30 604452, OGLE BLG-CEP-25, $P$=1.33916 days) and V1529 Sco (OGLE BUL-SC4 404186, OGLE BLG-CEP-14, OGLE DIA BUL-SC4 170, $P$=1.53185 days) we relay on the classification given in Soszy\'{n}ski et al. (2011b) stating that these stars are DCEPs. In the following we provide some details about some of the stars.


\end{multicols}

\noindent
\parbox{\textwidth}{
{\bf Table 3.} The identified DCEPs among short period BLHs in the GCVS.
\vskip.25cm \centerline{
\resizebox{\linewidth}{!}{
\begin{tabular}{l l l l l p{1.0cm} p{0.5cm} p{1.2cm} p{3.5cm} p{3.0cm}}
\hline\
Name & RA (h:m:s) & DEC (d:m:s) & $<$V$>$ [mag] &  $P$ [days] & $T_{\rm eff}$ [K] & $logg$ [cgs] & [Fe/H] [dex] & References & Notes \\
\hline
V351 Cep & 22:33:41.35 & +57:19:05.89 & 9.430 & 2.807  &  &  &  &  & \\
DQ And & 00:59:34.47 & +45:24:24.20 & 11.715 & 3.201 & 6596.19 & - & 0.418 & Luo et al. (2016) & \\
& & & & & 6366 5665 6068 & 2.82 1.93 1.79 & -0.08 -0.40 -0.02 & Schmidt et al. (2011) & \\
& & & & & 5500 & 1.5 & -0.5 & Maas et al. (2007) & \\
FM Del & 20:33:44.26 & +16:16:17.5 & 13.957 & 3.337 &  &  &  &  & Le Borgne and Klotz (2014): DCEP \\
BD Cas & 00:09:51.39 & +61:30:50.55 & 11.110 & 3.651 & 6200 6075 & 2.50 & -0.07 & Andrievsky et al. (2013) & \\
& & & & & - & - & -0.07 & Acharova et al. (2012) & \\
& & & & & 6069 6601 6278 & 2.40 4.70 1.72 & -0.03 0.02 -0.09 & Schmidt et al. (2011) & \\
& & & & & - & - & -0.07 & Luck et al. (2011) & \\
V572 Aql & 20:02:32.69 & +00:42:50.05 & 11.207 & 3.756 & 6250 & 1.0 & -0.2 & Maas et al. (2007) & \\
QY Cyg & 19:58:51.54 & +37:38:50.1 & 14.645 & 3.893 &  &  &  &  & Schmidt et al. (2005): DCEP \\
V383 Cyg & 20:28:58.16 & +34:08:06.36 & 10.909 & 4.612 &  &  &  &  &  \\
V675 Cen & 14:24:50.90 & -34:39:45 & 12.411 & 4.629 &  &  &  &  & Berdnikov et al. (2014):DCEP \\
V394 Cep & 22:02:40.90 & +59:27:09.04 & 14.008 & 5.688 &  &  &  &  & \\
AB Ara & 16:42:08.99 & -57:18:44.8 & 13.355 & 5.96 &  &  &  &  & ASAS, VSX, Fernie (1968) and Berdnikov et al. (2015): DCEP \\
TX Del & 20:50:12.69 & +03:39:08.36 & 9.158 & 6.166 & 6217 & 1.8 & 0.24 & Andrievsky et al. (2013) & \\ 
& & & & & - & - & 0.24 & Acharova et al. (2012) & \\
& & & & & 5738 5485 & 1.32 1.06 & 0.23 0.11 & Schmidt et al. (2011) & \\      
& & & & & - & - & 0.24 & Luck et al. (201) & \\
& & & & & 5500 & 0.5 & 0.1 & Maas et al. (2007) & \\
& & & & & 5553 5593 & - & 0.29 & Ammons et al. (2006) & \\
& & & & & 5900 & 1.6 & -0.18 &  Galazutdinov and Klochkova (1995) & \\
\hline
\end{tabular}
}  
}  
}
\vskip.5cm

\begin{multicols}{2}

	
	{\bf V351 Cep}
	
	V351 Cep ($P$=2.807 days, unclear) is a member of the galactic open cluster [KPR2005] (Zejda et al. 2012), which age was approximated to be 0.012 $\times$ 10$^9$ years. Galazutdinov and Klochkova (1995) from spectroscopic analysis, as well as taking into account previous photometric analysis especially from Arellano Ferro (1984), V351 Cep is not a DCEP, but it is not a T2C either, so they conclude that it is most probably an AC. However, Acharova et al. (2012) calculated the mass and age (as well as [Fe/H]=0.02) of V351 Cep to be 5.1~M$_{\odot}$ and 82~Myr old. Making this star too massive to be an AC. Balog et al. (1997) has suspected that V351 Cep is an s-Cepheid which pulsates in first overtone based on "small-amplitude sinusoidal light curve, radius as large as about 50-60~R$_{\odot}$, low galactic latitude".	
	
	The V-band data for the Fourier analysis was collected from Henden (1980) (19 data points) and Szabados (1977) (32 data points). The classification is not clear. It can be either a BLH or a DCEP, but if it is a BLH then all the other measurements (see above the mass and radius) would be incorrect. In conclusion, the true nature of V351 Cep is still open for discussion.
	
	
	{\bf DQ And}

	Szabados (1977) analysing light curve of DQ And ($P$=3.201 days) in V says that the presence of the bump in the descending branch and its high galactic latitude excludes the possibility that this is a DCEP. In the Schmidt et al. (2005) paper the V-band photometric observations do not show a bump on the descending branch, so the Szabados (1977) argument does not stand. However, Jurkovic et al. (2016) showed that this star in the span of 12 Gyrs has had a ellipsoidal orbit going from to bellow the Galactic plane. At this present moment it can be seen as a Halo object, but in the case of DQ And this does not mean that it is an old low-mass object. The position of DQ And on the $PR$ relation in Groenewegen and Jurkovic (2017b), using the radius from Balog et al. (1997): $R$=35 $\pm$ 6 R$_{\odot}$, as well as the derived Fourier parameters in this paper, from the  Schmidt et al. (2005) data, put it among the DCEPs.


	{\bf BD Cas} 

	 Andrievsky et al. (2013) have estimated the $T_{\rm eff}$ of BD Cas ($P$=3.651 days) to be 6200 and 6075 K. Acharova et al. (2012) derived the following element abundances from spectroscopic measurements: [O/H]=-0.09 and [Fe/H]=-0.07 [dex], from modelling calculated the age of BD~Cas to be 4.9 Myr, and the mass 4.9 M$_{\odot}$. In addition Schmidt et al. (2005) concluded, after much consideration, that BD Cas is an overtone pulsator. In our Fourier analysis of the INTEGRAL data we see that the parameters align with the DCEPs.


	{\bf TX Del, T2C or DCEP}

	Here we give an overview of the information available about TX Del ($P$=6.166 days), and while we would put is among DCEPs, we can not exclude the possibility that might be a T2C showing interesting features due to it being in a binary. 	
	Laney and Stobie (1995), Balog and Vinko (1995) and Balog et al. (1997) conclude from the Baade-Wesselink analysis that the radius of the star is too big for a T2C, one has to be careful with this assessment, again, because of the presence of a secondary component. Galazutdinov and Klochkova (1995) presented an extensive spectroscopic analysis and concluded that it shows all the hallmark of a classical Cepheid, but because of the binary nature and a possible interaction between the components. The automated classification of the {\it{Hipparcos}} variables (Dubath et al. 2011) has it as "DCEP". Acharova et al. (2012) has the mass estimate at $M$=6.4 M$_{\odot}$, and age 60 Myr, making it impossible to be T2C. Maas et al. (2007) points to Andrievsky et al. (2002) to say that this is a first-overtone Type I Cepheid, and it is clear from its chemical element abundance that it has gone through a mass transfer, nevertheless the authors stay with a conclusion that TX Del is a T2C. 


\section{5. STARS THAT ARE NOT PULSATING VARIABLES IN THE SAMPLE}
\label{sec:other}

	{\bf NY Her}
	
	NY Her ($P$=0.076 days) is a SU UMa type dwarf novae as it was described in Kato et al. (2013). We confirm their findings on the AAVSO dataset of NY Her as shown on Fig. 9a.


	{\bf V4110 Sgr} 

	Fig. 9b shows our analysis of the OGLE I-band data of V4110 Sgr ($P$=1.125 days). The period that we found was $P$=1.1254 days. It has a strange jump in amplitude, which comes back to the previous level. It would be highly unlikely for a pulsating star to show such a behaviour, if it was real. This feature, on the other hand, could be a result of a measurement error, so we can not give a definite conclusion. Another thing to notice is that V4110 Sgr is flagged to be in a crowded field, and a neighbouring star OGLEII DIA BUL-SC01 V3246 (RA: 270.757455, DEC: -29.766070, and the median I-band magnitude is 15.899 $\pm$ 0.047 mag) is a BY Draconis variables type of variable star (rotating stars with star spots, and other chromospheric activity).
	Blanco (1984) has V4110 Sgr observed from 1977 to 1981 (it was originally classified as a CW star by Gaposchkin), and it showed a variable period (in 1977 $P$=1.1247, in 1979 $P$=1.1345, and in 1980 $P$=1.108 days).


	{\bf CT Sge}
	
	CT Sge ($P_{GCVS}$=1.7179 days) has V-band data in ASAS catalogue, but the period given by the GCVS team, $P$=1.7179 days, and the period from the Fourier analysis does not phase the light curve. We were not able to phase the light curve using the the above periods, nor could we establish a new one from the Fourier analysis.

	
	{\bf DI And}
	
	DI And ($P_{GCVS}$=3.385584882 days) was originally classified in the GCVS as an "IS" star, which changed to "CWB:" after Khruslov (2005) published it's results, noting that "the shape of the descending branch is not quite typical". WASP has a long data set, which we looked at (see Fig. 9c). We are not aware of any known T2C nor DCEP of this period to have an even remotely similar light curve shape. Morrison et al. (2001) gives us the information that it is not a binary. The possible types of variability that could result in a similar light curve shape could be a star with spots, a chromospherically active star or a rotating star.


	{\bf UW For}
	
	UW For ($P_{\rm ASAS}$=4.277943728 days) is an eclipsing binary system, as it was described by the ASAS with a period of $P$=4.27774 days.

	The phased light curve (see Fig. 9d) shows the two minima in the eclipse. We do not exclude the possibility that after removing the eclipse from the light curve there could be other variability left in the system.
	

	{\bf KT Com}
	
	KT Com ($P_{\rm ASAS}$=8.140512308 days) is listed as a semi-detached (ESD) or detached (ED) binary system. Using this classification  Szczygiel et al. (2008) has calculated the bolometric luminosity $L$=33.418 [ergs/sec], X-ray luminosity $L_{\rm X-ray}$=29.765 [ergs/sec], and distance, d=39.1256 [pc]. Independently Soubiran et al. (2016) gives an effective temperature $T_{\rm eff}$=5924~K for this star.
	In the "The International Variable Star Index" (VSX) of the The American Association of Variable Star Observers ("AAVSO")\footnote{http://www.aavso.org/vsx/} KT Com is a variable star of RS Canum Venaticorum-type binary system, with a period of P=4.07 days, but the given reference, Kiraga (2012), lists this star as a rotation variable.
	We conclude that KT Com is not a BLH, but instead it is most probably a binary system.


	{\bf V403 Cyg (no period)}
	
	Koch (1974) listed V403 Cyg as a strong interacting binary, but there is no light curve that would support this claim. The GCVS listed Suzuki and Huruhata (1938) for the identification, and it is listed as a binary in Coughlin et al. (2014), yet again there were no data to support it. SuperWASP has a lot of data points, but it does not show any sign of binarity and it can not be phased with the given period of P$_{\rm GCVS}$=0.80477 days. This is too short for a BLH, but it is in the range of ACs. Taken into account that Hanson et al. (2004) has classified it as an RRL, it is very possible that indeed it is an AC, but further observations would be needed to confirm this claim. We could not determine what kind of variable this star is. 
	
\end{multicols}
\begin{minipage}[b]{.4\linewidth}
\centering
\includegraphics[width=0.9\columnwidth, keepaspectratio]{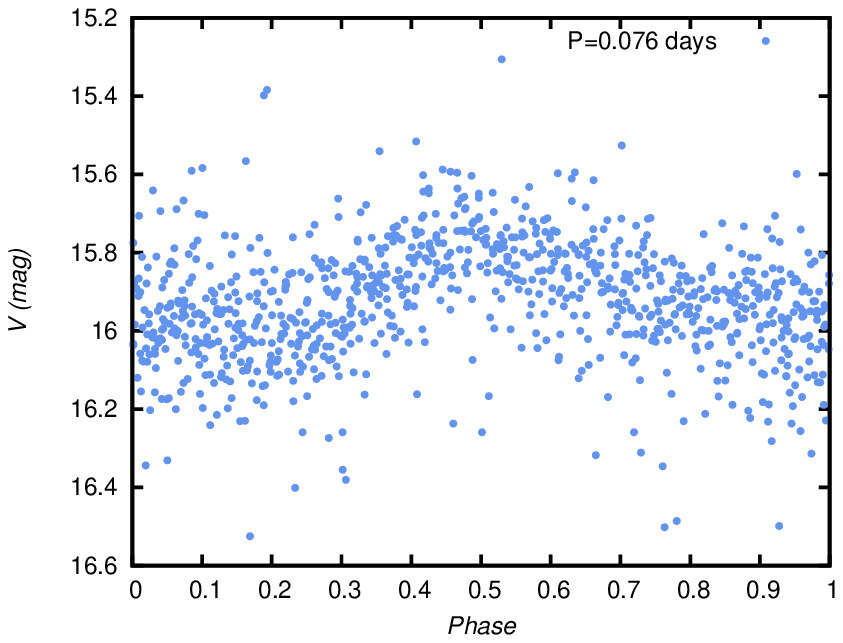}
\\
{(a) NY Her phased with $P$=0.076 days from the AAVSO data.} 
\end{minipage}
\begin{minipage}[b]{.4\linewidth}
\centering
{\includegraphics[width=0.9\columnwidth, keepaspectratio]{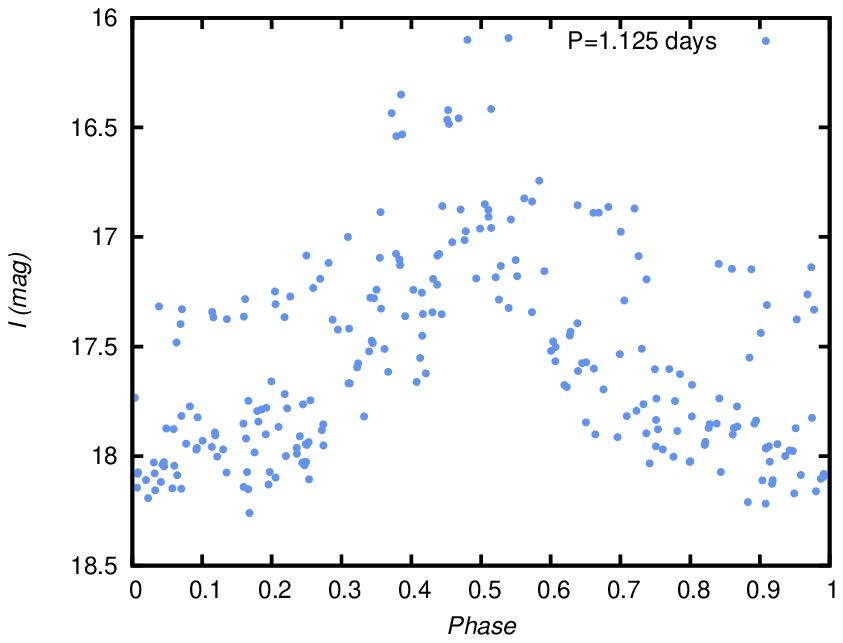}}
\\
{(b) V4110 Sgr OGLE II I-band phased light curve with $P$=1.125 days.} 
\end{minipage}

\begin{minipage}[b]{.4\linewidth}
\centering
\includegraphics[width=0.9\columnwidth, keepaspectratio]{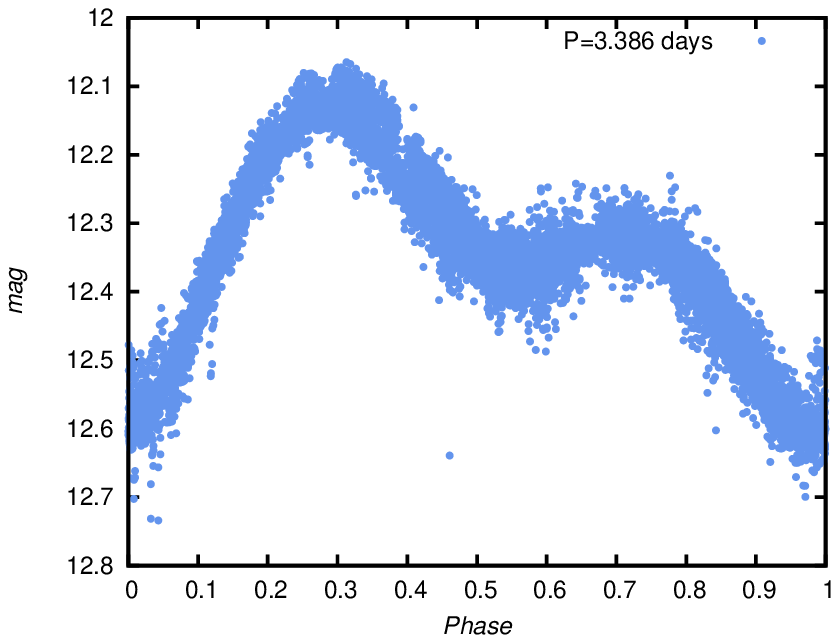}
\\
{(c) DI And phased light curve from SuperWASP data with $P$=3.386 days.} 
\end{minipage}
\begin{minipage}[b]{.4\linewidth}
\centering
{\includegraphics[width=0.9\columnwidth, keepaspectratio]{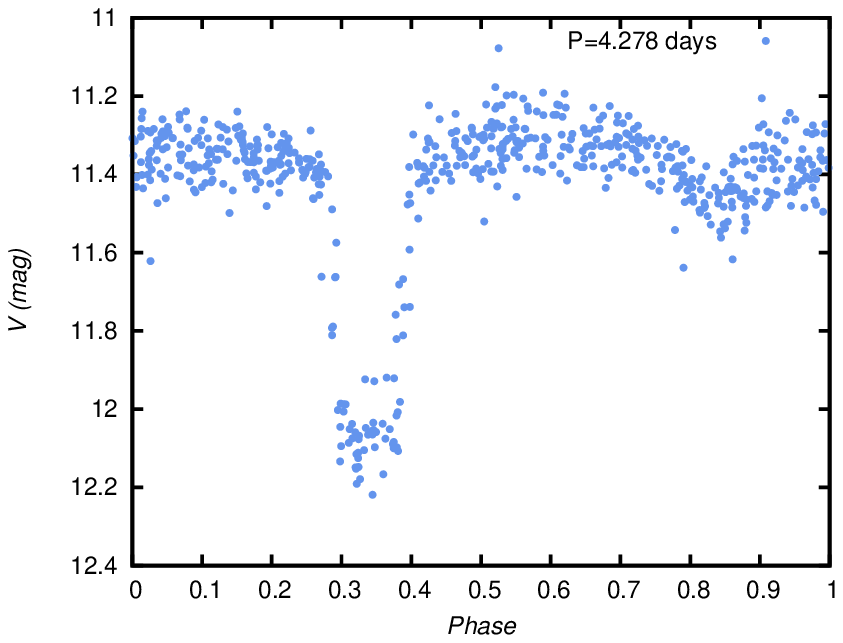}}
\\
{(d) UW For phased light curve with $P$=4.278 days using ASAS-3 data.} 
\end{minipage}

\begin{minipage}[b]{.4\linewidth}
\centering
\includegraphics[width=0.9\columnwidth, keepaspectratio]{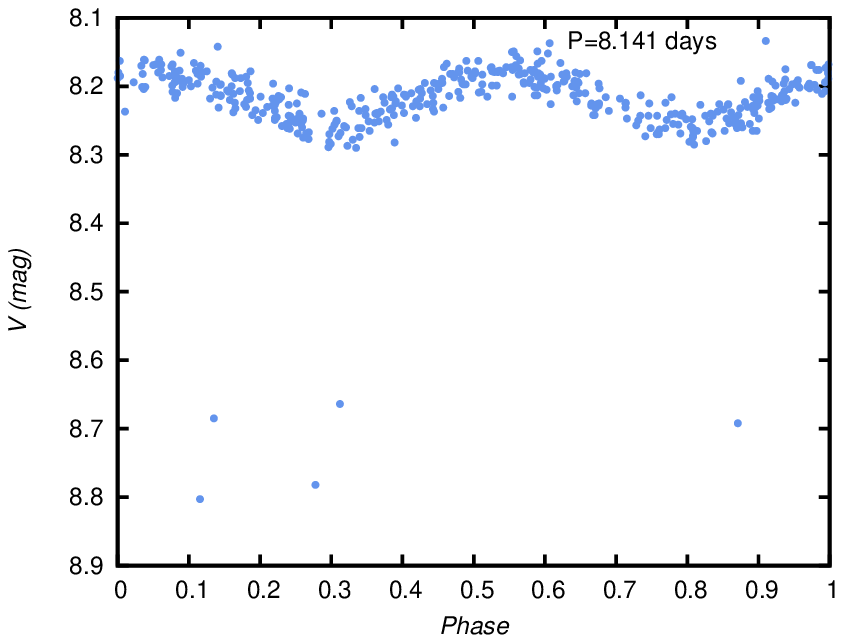}
\\
{(e) KT Com phased light curve with a period $P$=8.141 days using ASAS-3 data.} 
\end{minipage}

\figurecaption{9.}{Stars that are not pulsating variables.}

\begin{multicols}{2}	

\section{6. THE DISCUSSION OF METALLICITY}
\label{sec:metallicity}

	The collected spectroscopic data of the published $T_{\rm eff}$, log g, [Fe/H] confirms a result from Diethelm (1990), described in Wallerstein (2002) about the metallicities. Diethelm (1990) has estimated the photometric metallicities of the than known short period T2Cs, and it became obvious that from the 45 stars in his list, 30 had [Fe/H]$_{\rm VBLU} >$ -0.3, and 8 had [Fe/H]$_{\rm VBLU} <$ -1.0. Comparing our list of AC with these results we saw that 7 out of 8 were indeed AC. The metallicities that are in Tables 1 and 2 are directly measured from spectra, and they also show the same distribution of [Fe/H]. In the case of the ACs we have only data for four stars, and they are very metal poor. The average of all the eight measurements of four stars is $<$[Fe/H]$>$=-1.12 dex, but if we leave out the outlier of UY Eri by Schmidt et al. (2011) then the it becomes $<$[Fe/H]$>$=-1.88 dex. While the number of measured metallicities is small, we do suggest that it is correct to conclude that this low metallicity is fundamental to the evolution of these objects. This was already known from observations in dwarf spherical and dwarf irregular galaxies (for example Sculptor, Sextans, LeoII, Ursa Minor, Draco, Fornax, LeoI) and in smaller numbers in globular clusters, such as BL Boo in NGC 5466, and in the Milky Way summarized in Fiorentino and Monelli (2012). Bono (1997b) and Fiorentino and Monelli (2012), as well as others, have used these observational boundaries to constrain their theoretical calculations confirming that ACs are metal-poor stars with an mean mass of 1.2 - 1.5 M$_{\odot}$.
		
	The BLH stars, on the other hand, turned out to have Solar metallicities on average, $<$[Fe/H]$>$=0.00 dex. We got this by combining all the twenty one measurements of ten stars. BLHs are not metal poor, despite being old objects.
	
	In order to see how these result reflect on the Herztsprung-Russell diagram (HRD) of the BLH stars in the MW we have calculated the luminosities (L) from the Groenewegen and Jurkovic (2017b). We used the $M_{bol}$=0.12-1.78*logP formula to calculate the bolometric magnitudes, and than we converted that to luminosities. The T$_{eff}$ values are given in Table 2, which were measured from spectroscopy. In the cases were there were multiple measurements for $T_{\rm eff}$ we have used the mean value of all the values. In the same table we give the values of pulsational periods in days, derived from our Fourier analysis. The number of BLH stars was limited to ten stars, since only for these were the spectroscopy was available, and we give all the relevant data in Table 4. In Fig. 10 we have compared our computed $L$ and $T_{\rm eff}$ with models of horizontal branch (HB) stars from the BaSTI\footnote{http://basti.oa-teramo.inaf.it/} evolutionary model database. The models were calculated for the Solar metallicity of [M/H]=0.058 to match our findings, and previously assumed metallicity value for metal-poor stars of [M/H]=-1.488. The masses of the modelled stars were $M$=0.500 M$_{\odot}$, $M$=0.520 M$_{\odot}$, $M$=0.530 M$_{\odot}$ and $M$=0.550 M$_{\odot}$. These mass ranges are in agreement with estimates from Bono (1997a) of approx. 0.52-0.53 M$_{\odot}$ and Groenewegen and Jurkovic (2017a) of 0.49 M$_{\odot}$. All of our stars fall in the region of Solar metallicity models, with the majority being on the track for the  $M$=0.500 M$_{\odot}$ model, and going towards the track of the $M$=0.520 M$_{\odot}$ model. The metal-poor model of $M$=0.550 M$_{\odot}$ is very close to the Solar metallicity model of $M$=0.500 M$_{\odot}$, but the independently measured spectroscopic metallicities as well as the newest mass estimates seem to strengthen our finding that BLH stars have Solar-like metallicities.


\end{multicols}

\noindent
\parbox{\textwidth}{
{\bf Table 4.} The mean effective temperature ($<T_{\rm eff}>$) from the spectroscopic measurements shown in Table 2, the pulsational period ($P$), the bolometric magnitude form $M_{bol}$=0.12-1.78*logP (Groenewegen and Jurkovic 2017b), and the calculated luminosities ($L$) for the Galactic BLH stars.
\vskip.25cm \centerline{
\begin{tabular}{llllllll}
\hline\
Name &  $<T_{\rm eff}>$ (K) & log$<T_{\rm eff}>$ (K) & $P$ (days) & log$P$ (days) & $M_{bol}$ (mag) & $L$ (L$_{\odot}$) & log$L$ (L$_{\odot}$) \\
\hline
BX Del & 6250 & 3.796  & 1.092 & 0.038 & 	0.052 & 75.03 & 1.87\\
VY Pyx & 5750 & 	3.760 & 1.240 & 0.093 & -0.046 & 82.13 & 1.91\\
V527 Sgr & 5816 & 3.765 & 1.255 & 0.099 & -0.056 & 82.84 & 1.92\\
KZ Cen & 6141 & 	3.788 & 1.520 & 0.182 & -0.204 & 94.94 & 1.98\\
SW Tau & 6772.5 & 3.831 & 1.584 & 0.200 & -0.236 & 97.77 & 1.99\\
V439 Oph & 5547 & 3.744 & 1.893 & 0.277 & -0.373 & 111.00 & 2.05\\
V1287 Sco & 	5428	 & 3.735 & 2.036 & 	0.309 & -0.430 & 116.91 & 2.08\\
V553 Cen & 5629.7 & 3.750 & 2.061 & 0.314 & -0.440 & 117.93 & 2.07\\
RT TrA & 6092.8 & 3.785 & 1.946 & 0.289 & -0.395 & 113.21 & 2.05\\
BL Her & 6364.7 & 3.804 & 1.307 & 0.116 & -0.087 & 85.27 & 1.931\\
\hline
\end{tabular}
}  
}  
\vskip.5cm


\centerline{\includegraphics[width=\columnwidth]{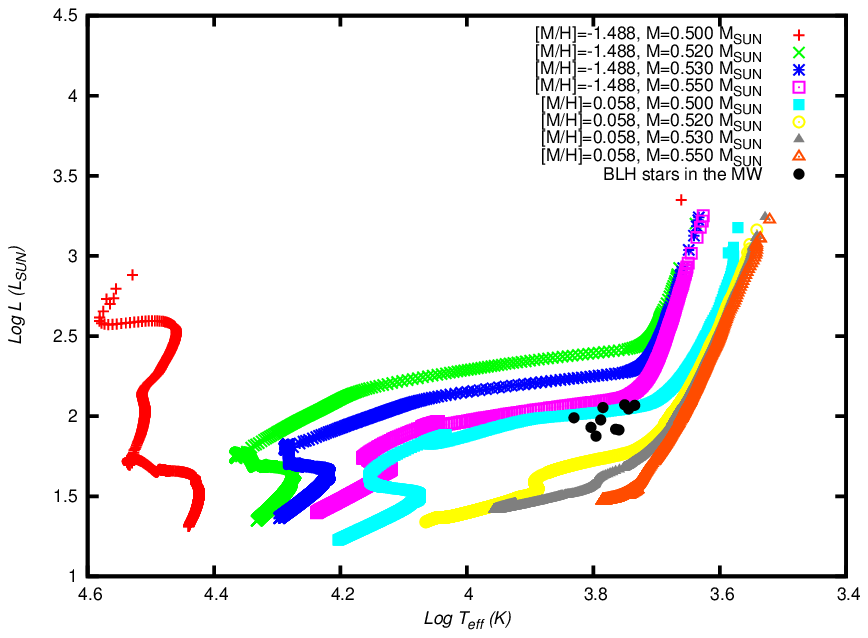}}
\figurecaption{10.}{The BLH stars from the MW plotted with a metal-poor and a Solar metallicity models. The models are from the BaSTI database (Pietrinferni et al. 2004). The black circles represent the data for BLH stars shown in Table 4. In the case on the metal-poor models ([M/H]=-1.488) the masses of are shown with the following colors and symbols: red plus signs $M$=0.500 M$_{\odot}$, green crosses $M$=0.520 M$_{\odot}$, blue star $M$=0.530 M$_{\odot}$ and pink doted square $M$=0.550 M$_{\odot}$. For the Solar metallicity models ([M/H]-0.058) the masses are shown as: cyan square $M$=0.500 M$_{\odot}$, yellow dotted circle $M$=0.520 M$_{\odot}$, grey filled triangle $M$=0.530 M$_{\odot}$, orange dotted triangle $M$=0.550 M$_{\odot}$.}

\begin{multicols}{2}

%

\section{7. CONCLUSIONS}
\label{sec:conclusions}

	Fourier parameters of 59 stars from the GCVS short period T2Cs list, known as BLHs, as well as 33 BLHs and 29 ACs from the LMC OGLE-III catalogue for comparison, using V-band datasets, were derived. Among the 59 stars we have found 19 F ACs, 1 1O AC, 26 BLHs, 2 pWVir/DCEP, 11 DCEPs, and 7 other types of variables.
	
	The BLHs and ACs can be distinguished using the $R_{21}$, $R_{31}$ and $\phi_{31}$ Fourier parameters. The $\phi_{21}$ parameter is not useful fr his purpose, because almost all the points clumped together. Interestingly, Figs. 2 and 3, representing the amplitude of the two and three times of the fundamental frequency, showed a very clear separation between the ACs and the BLHs. These Fourier parameters can be used as new comparison sample for the stars measured by the Kepler space telescope, since its broad band filter corresponds broadly with the V-band.
		
	In total 19 F ACs and one 1O AC was discovered in the examined sample. FY Vir, V563 Cen, V716 Oph, BF Ser, BI Tel, VX Cap, XX Vir and V1149 Her were detected as ACs in the CSS (Drake et al. 2014a, 2017) using a different method of detection, and we confirmed their classification through the Fourier parameters. We add further 12 F ACs (FY Aqr, PP Tel, DF Hyi, BQ CrA, BH Cet, V2733 Oph, CE Her, MQ Aql, V745 Oph, UY Eri, UX Nor), and one 1O AC (V742 Cyg) to the previously known once. We list some of our findings:

\item{(i)} V742 Cyg is a 1O AC, and a supposed member of Dolidze 37 open cluster, making it a first objects of this type to be discovered in an open cluster, but further observations would be needed to confirm this result;
\item{(ii)} FY Vir shows signs of Blazhko effect-like modulation.

	BLH star are shown to have Solar-like metallicities if the AC are properly separated from the sample. The AC, despite being more massive, seem to have low-metallicities.
	
	These results can be useful for other research fields as well, such as Galactic archaeology and evolution modelling, as well for individual measurements of these objects, and for the improvement of the precision of the $PL$ relation, as it will become even more important with the new Gaia data (Gaia Collaboration 2016).
	


\acknowledgements{
I would like to express my deepest gratitude to G. B., Martin Groenewegen, R\'{o}bert Szab\'{o}, Emese Plachy, Branislav Vukoti\'{c}, L\'{a}szl\'{o} Moln\'{a}r, and all the colleagues with whom I have consulted for all of their help and patience. This research had financial support from the Ministry of Education, Science and Technological Development of the Republic of Serbia through the project 176004. This project has been supported by the NKFIH K-115709 grant of the Hungarian National Research, Development and Innovation Office.
	This research has made use of NASA's Astrophysics Data System.
	This research has made use of the SIMBAD database,
operated at CDS, Strasbourg, France (Wenger et al. 2000).
	Based on data from the OMC Archive at CAB (INTA-CSIC), pre-processed by ISDC.
	The CSS survey is funded by the National Aeronautics and Space
Administration under Grant No. NNG05GF22G issued through the Science
Mission Directorate Near-Earth Objects Observations Program.  The CRTS
survey is supported by the U.S.~National Science Foundation under
grants AST-0909182.
	This research has made use of the NASA/ IPAC Infrared Science Archive, which is operated by the Jet Propulsion Laboratory, California Institute of Technology, under contract with the National Aeronautics and Space Administration.
	This work has made use of data from the European Space Agency (ESA)
mission Gaia (https://www.cosmos.esa.int/gaia), processed by
the Gaia Data Processing and Analysis Consortium (DPAC,
https://www.cosmos.esa.int/web/gaia/dpac/consor \break tium). Funding
for the DPAC has been provided by national institutions, in particular
the institutions participating in the Gaia Multilateral Agreement.
}


\references

Acharova, I. A., Mishurov, Y. N., and Kovtyukh, V. V.: 2012, \journal{Mon. Not. R. Astron. Soc.}, \vol{420}, 1590.

Allende Prieto, C., Rebolo, R., Garc\'{i}a L\'{o}pez, R. J., et al.: 2000, \journal{Astron. J.}, \vol{120}, 1516.

Ammons, S. M., Robinson, S. E., Strader, J., et al.: 2006, \journal{Astrophys. J.}, \vol{638}, 1004.

Andrievsky, S. M., Kovtyukh, V. V., Luck, R. E., et al.: 2002, \journal{Astron. Astrophys}, \vol{381}, 32.

Andrievsky, S. M., L\`{e}pine, J. R. D., Korotin, S. A., et al.: 2013, \journal{Mon. Not. R. Astron. Soc.}, \vol{428}, 3252.

Arellano Ferro, A.: 1984, \journal{Mon. Not. R. Astron. Soc.}, \vol{209}, 481.

Arellano Ferro, A. and Rosenzweig, P.: 2000, \journal{Mon. Not. R. Astron. Soc.}, \vol{315}, 296.

Balog, Z. and Vinko, J.: 1995, \journal{Inf. Bull. Var. Stars}, \vol{4150}.

Balog, Z., Vinko, J., and Kaszas, G.: 1997, \journal{Astron. J.}, \vol{113}, 1833.

Berdnikov, L. N.: 2008, \textit{VizieR Online Data Catalog}, 2285: Photoelectric observations of Cepheids in UBV(RI)c.

Berdnikov, L. N., Kniazev, A. Y., Sefako, R., Kravtsov, V. V., and Zhujko, S. V.: 2014, \journal{Astron. Lett.}, \vol{40}, 125.

Berdnikov, L. N., Kniazev, A. Y., Sefako, R., et al.: 2015, \journal{Astron. Lett.}, \vol{41}, 23.

Blanco, B. M.: 1984, \journal{Astron. J.}, \vol{89}, 1836.

Bobylev, V. V.: 2008, \journal{Astron. Lett.}, 34, 686.

Bono, G., Caputo, F., and Santolamazza, P.: 1997a,  \journal{Astron. Astrophys}, \vol{317}, 171.

Bono, G., Caputo, F., Santolamazza, P., Cassisi, S., and Piersimoni, A.: 1997b, \journal{Astron. J.}, \vol{113}, 2209.

Caldwell, C. N. and Butler, D.: 1978, \journal{Astron. J.}, \vol{83}, 1190.

Catelan, M. and Smith, H. A.: 2015, Pulsating Stars, Wiley-VCH Verlag GmbH and Co., Weinheim.

Clement, C. M., Muzzin, A., Dufton, Q., et al.: 2001, \journal{Astron. J.}, \vol{122}, 2587.

Clement, C. M. 2017, \textit{VizieR Online Data Catalog}, 5150: Updated catalog of variable stars in globular clusters.

Coughlin, J. L., Thompson, S. E., Bryson, S. T., et al.: 2014, \journal{Astron. J.}, \vol{147}, 119.

Diethelm, R.: 1983, \journal{Astron. Astrophys}, \vol{124}, 108.

Diethelm, R.: 1990, \journal{Astron. Astrophys}, \vol{239}, 186.

Dinescu, D. I.: 2002, \journal{ASP Conf. Ser.}, \vol{265}, 365.

Drake, A. J., Djorgovski, S. G., Mahabal, A., et al.: 2009, \journal{Astrophys. J.}, \vol{696}, 870.

Drake, A. J., Graham, M. J., Djorgovski, S. G., et al.: 2014a, \journal{Astrophys. J. Suppl. Ser.}, \vol{213}, 9.

Drake, A. J., Graham, M. J., Djorgovski, S. G., et al.: 2014b, \textit{VizieR Online Data Catalog}, J/ApJS/213/9: Catalina Surveys periodic variable stars.

Drake, A. J., Djorgovski, S. G., Catelan, M., et al.: 2017, \journal{Mon. Not. R. Astron. Soc.}, \vol{469}, 3688.

Dubath, P., Rimoldini, L., Süveges, M., et al.: 2011, \journal{Mon. Not. R. Astron. Soc.}, \vol{414}, 2602.

Fernie, J. D.: 1968, \journal{Astron. J.}, \vol{73}, 995.

Fiorentino, G., Limongi, M., Caputo, F., and Marconi, M.: 2006, \journal{Astron. Astrophys}, \vol{460}, 155.

Fiorentino, G. and Monelli, M.: 2012, \journal{Astron. Astrophys}, \vol{540}, A102.

Fokin, A. B. and Gillet, D.: 1994, \journal{Astron. Astrophys}, \vol{290}, 875.

Gaia Collaboration, Prusti, T., de Bruijne, J. H. J., et al.: 2016, \journal{Astron. Astrophys}, \vol{595}, A1.

Galazutdinov, G. A. and Klochkova, V. G.: 1995, \journal{Astron. Astrophys. Trans.}, \vol{8}, 227.

Gillet, D., Burki, G., Chatel, A., Duquennoy, A., and Lebre, A.: 1994, \journal{Astron. Astrophys}, \vol{286}, 508.

Groenewegen, M. A. T. and Jurkovic, M. I.: 2017a, \journal{Astron. Astrophys}, \vol{603}, A70.

Groenewegen, M. A. T. and Jurkovic, M. I.: 2017b, \journal{Astron. Astrophys}, \vol{604}, A29.

Hanson, R. B., Klemola, A. R., Jones, B. F., and Monet, D. G.: 2004, \journal{Astron. J.}, \vol{128}, 1430.

Harris, H. C.: 1981, \journal{Astron. J.}, \vol{86}, 719.

Henden, A. A.: 1980, \journal{Mon. Not. R. Astron. Soc.}, \vol{192}, 621.

Huber, D., Bryson, S. T., Haas, M. R., et al.: 2016, \journal{Astrophys. J. Suppl. Ser.}, \vol{224}, 2.

Jurkovi\'{c}, M. and Szabados, L.: 2014, \journal{IAU Symp.}, \vol{301}, 431–432.

Jurkovic, M. I.: 2015, \textit{European Physical Journal Web of Conferences}, \vol{101}, 06035.

Jurkovic, M. I., Stojanovic, M., and Ninkovic, S.: 2016, Commmunications of the Konkoly Observatory Hungary, \vol{105}, 175.

Kato, T., Hambsch, F.-J., Maehara, H., et al.: 2013, \journal{Publ. Astron. Soc. Jpn.}, \vol{65}, 23.

Khruslov, A. V.: 2005, \textit{Peremennye Zvezdy Prilozhenie}, \vol{5}.

Kiraga, M.: 2012, \journal{Acta Astron.}, \vol{62}, 67.

Koch, R. H.: 1974, \journal{Astron. J.}, \vol{79}, 34.

Kwee, K. K. and Diethelm, R.: 1984, \journal{Astron. Astrophys}, \vol{55}, 77.

Laney, C. D. and Stobie, R. S.: 1995, \journal{Mon. Not. R. Astron. Soc.}, \vol{274}, 337.

Layden, A. C.: 1994, \journal{Astron. J.}, \vol{108}, 1016.

Le Borgne, J. F. and Klotz, A.: 2014, arXiv:1407.4961.

Lenz, P. and Breger, M.: 2004, \journal{IAU Symp.}, \vol{224}, 786–790.

Lloyd Evans, T.: 1983, \journal{Observatory}, \vol{103}, 276.

Luck, R. E., Andrievsky, S. M., Kovtyukh, V. V., Gieren, W., and Graczyk, D.: 2011, \journal{Astron. J.}, \vol{142}, 51.

Luo, A.-L., Zhao, Y.-H., Zhao, G., et al.: 2016, \textit{VizieR Online Data Catalog}, 5149: LAMOST DR2 catalogs.

Maas, T., Giridhar, S., and Lambert, D. L.: 2007, \journal{Astrophys. J.}, \vol{666}, 378.

Madore, B. F. and Fernie, J. D.: 1980, \journal{Publ. Astron. Soc. Pac.}, \vol{92}, 315.

McCarthy, J. K. and Nemec, J. M.: 1997, \journal{Astrophys. J.}, \vol{482}, 203.

McDonald, I., Zijlstra, A. A., and Boyer, M. L.: 2012, \journal{Mon. Not. R. Astron. Soc.}, \vol{427}, 343.

Morrison, J. E., R\"{o}ser, S., McLean, B., Bucciarelli, B., and Lasker, B.: 2001, \journal{Astron. J.}, \vol{121}, 1752.

Mu\~{n}oz Bermejo, J., Asensio Ramos, A., and Allende Prieto, C.: 2013, \journal{Astron. Astrophys}, \vol{553}, A95.

Nemec, J. and McCarthy, J. K.: 1998, \journal{ASP Conf. Ser.}, \vol{135}, 57.

Petersen, J. O. and Diethelm, R.: 1986, \journal{Astron. Astrophys}, \vol{156}, 337.

Pojmanski, G.: 1997, \journal{Acta Astron.}, \vol{47}, 467.

Pojmanski, G., Pilecki, B., and Szczygiel, D.: 2005, \journal{Acta Astron.}, \vol{55}, 275.

Provencal, J.: 1986, \textit{Journal of the American Association of Variable Star Observers}, \vol{15}, 36.

Richards, J. W., Starr, D. L., Miller, A. A., et al.: 2012, \journal{Astrophys. J. Suppl. Ser.}, \vol{203}, 32.

Samus, N. N., Durlevich, O. V., et al.: 2009a, \textit{VizieR Online Data Catalog}, J/PASP/121/1378: Positions of variable stars in globular clusters.

Samus, N. N., Kazarovets, E. V., Pastukhova, E. N., Tsvetkova, T. M., and Durlevich, O. V.: 2009b, \journal{Publ. Astron. Soc. Pac.}, \vol{121}, 1378.

Samus, N. N., Kazarovets, E. V., Durlevich, O. V., Kireeva, N. N., and Pastukhova, E. N.: 2017, \journal{Astron. Rep.}, \vol{61}, 80.

Schmidt, E. G., Loomis, C. G., Groebner, A. T., and Potter, C. T.: 1990, \journal{Astrophys. J.}, \vol{360}, 604.

Schmidt, E. G.: 1991, \journal{Astron. J.}, \vol{102}, 1766.

Schmidt, E. G. and Reiswig, D. E.: 1993, \journal{Astron. J.}, \vol{106}, 2429.

Schmidt, E. G., Lee, K. M., Johnston, D., Newman, P. R., and Snedden, S. A.: 2003, \journal{Astron. J.}, \vol{126}, 906.

Schmidt, E. G., Johnston, D., Langan, S., and Lee, K. M.: 2005, \journal{Astron. J.}, \vol{130}, 832.

Schmidt, E. G., Rogalla, D., and Thacker-Lynn, L.: 2011, \journal{Astron. J.}, \vol{141}, 53.

Sharma, S., Bland-Hawthorn, J., Johnston, K. V., and Binney, J.: 2011, \journal{Astrophys. J.}, \vol{730}, 3.

Simon, N. R. and Lee, A. S.: 1981, \journal{Astrophys. J.}, \vol{248}, 291.

Soszy\'{n}ski, I., Udalski, A., Szyma\'{n}ski, M. K., et al.: 2008, \journal{Acta Astron.}, \vol{58}, 293.

Soszy\'{n}ski, I., Poleski, R., Udalski, A., et al.: 2010a, \journal{Acta Astron.}, \vol{60}, 17.

Soszy\'{n}ski, I., Udalski, A., Szyma\'{n}ski, M. K., et al.: 2010b, \journal{Acta Astron.}, \vol{60}, 91.

Soszy\'{n}ski, I., Dziembowski, W. A., Udalski, A., et al.: 2011a, \journal{Acta Astron.}, \vol{61}, 1.

Soszy\'{n}ski, I., Udalski, A., Pietrukowicz, P., et al.: 2011b, \journal{Acta Astron.}, \vol{61}, 285.

Soszy\'{n}ski, I., Udalski, A., Szyma\'{n}ski, M. K., et al.: 2017, arXiv:1712.01307.

Soubiran, C., Le Campion, J.-F., Brouillet, N., and Chemin, L.: 2016, \journal{Astron. Astrophys}, \vol{591}, A118.

Suzuki, K. and Huruhata, M.: 1938, \journal{Astron. Nachr.}, \vol{267}, 101.

Szabados, L.: 1977, \textit{Commmunications of the Konkoly Observatory Hungary}, \vol{70}, 1.

Szabados, L., Kiss, L. L., and Derekas, A.: 2007, \journal{Astron. Astrophys}, 461, 613.

Szczygie\l{}, D. M., Socrates, A., Paczy\'{n}ski, B., Pojma\'{n}ski, G., and Pilecki, B.: 2008, \journal{Acta Astron.}, \vol{58}, 405.

Udalski, A., Kubiak, M., Szymanski, M., et al.: 1994, \journal{Acta Astron.}, \vol{44}, 317.

Udalski, A.: 1996, \textit{VizieR Online Data Catalog}, II/213/pvar: OGLE Galactic Bulge periodic variables.

Vasil’Yanovskaya, O. P.: 1978, \journal{Peremennye Zvezdy}, \vol{21}, 111.

Vinko, J., Remage Evans, N., Kiss, L. L., and Szabados, L.: 1998, \journal{Mon. Not. R. Astron. Soc.}, \vol{296}, 824.

Wallerstein, G., Brown, J. A., and Bates, B. A.: 1979, \journal{Publ. Astron. Soc. Pac.}, \vol{91}, 47.

Wallerstein, G. and Gonzalez, G.: 1996, \journal{Mon. Not. R. Astron. Soc.}, \vol{282}, 1236.

Wallerstein, G., Matt, S., and Gonzalez, G.: 2000, \journal{Mon. Not. R. Astron. Soc.}, \vol{311}, 414.

Wallerstein, G.: 2002, \journal{Publ. Astron. Soc. Pac.}, \vol{114}, 689.

Wenger, M., Ochsenbein, F., Egret, D., et al.: 2000, \journal{Astron. Astrophys Suppl. Ser.}, \vol{143}, 9.

Zejda, M., Paunzen, E., Baumann, B., Mikul\'{a}\v{s}ek, Z., and Li\v{s}ka, J.: 2012, \journal{Astron. Astrophys}, \vol{548}, A97.

Zinn, R. and Dahn, C. C.: 1976, \journal{Astron. J.}, \vol{81}, 527.

\endreferences

\end{multicols}

\vfill\eject

{\ }



\naslov{ANOMALNE CEFEIDE OTKRIVENE U UZORKU GALAKTIQKIH KRATKOPERIODIQNIH CEFEIDA TIPA $\mathbf{II}$}


\authors{M. I. Jurkovic$^{1,2}$}
\vskip3mm

\address{$^1$Astronomical Observatory, Volgina 7, 11060 Belgrade, Serbia}

\Email{mojur}{aob.rs}

\address{$^2$Konkoly Thege Mikl\'{o}s Astronomical Institute, Research Centre for Astronomy and Earth Sciences, Hungarian Academy of Sciences\break H-1121 Budapest, Konkoly Thege Mikl\'{o}s \'{u}t 15-17.}

\vskip.7cm


\centerline{UDK \udc}


\centerline{\rit Uredjivaqki prilog}

\vskip.7cm

\begin{multicols}{2}
{


{\rrm Ponovo smo posetili kratkoperiodiqne cefeide tipa} II{\rrm, zvane} BL Herculis {\rrm (BLHs), u Galatiqkom polju da bismo izveli homogenu analizu njihovih Furije parametara.
	
	Koristili smo iskljuqivo podatke sakupljene u} V {\rrm filtru da bismo obezbedili da se promeljive iz} OGLE-III {\rrm kataloga mogu direktno uporediti sa 59 zvezda, koje su bile klasifikovane kao kratkoperiodiqne cefeide tipa} II {\rrm u Generalnom Katalogu Prome{\lj}ivih Zvezda u naxem uzorku.
	
	Od 59 zvezda naxli smo da je 19 BLH tip, 19 su anomalne cefeide (ACs) koje pulsiraju u fundamentalnoj modi (8 od ovih je ve{\cc} prethodno bilo klasifikovano od strane Katalina Nebeskog Pregleda, 1 je anomalna cefeida koja pulsira u prvoj nad modi, za 2 prepostavljamo da su naroqitog} W Virginis {\rrm tipa (pVVVir)), 11 su klasiqne cefeide (DCEPs), a 7 nisu pulsairaju{\cc}e zvezde uopxte. Kreirali smo listu sjajnih BLH zvezda u Galaksiji i razdvojili smo ACs, kao i druge objekte, koji su bili pogreqno klasifikovani. Broj stvarnih BLH zdezda u naqem uzorku se smanjio za viqe od 50\%. Sakupili smo metaliqnosti dobijene iz spektroskopskih merenja koji su bili publikovani u literaturi. Iako je broj stvarnih merenja mali, veoma je verovatno da su ACs niske metaliqnosti. Proseqna metaliqnost iz 8 merenja 4 zvezda (}UY Eri {\rrm ima 5 razliqitih merenja} [Fe/H]{\rrm ) je $-1.12$} dex{\rrm , ali ako izostavimo vrednosti metaliqnosti koje su drastiqno razlikuju kod} UY Eri {\rrm zvezde onda proseqna metaliqnost postaje $-1.88$} dex{\rrm , bez obzira na to da li se AC nalazi unutar samog Mleqnog puta ili u nekom od obliznjih jata. Sa druge strane, qini se da BLH zvezde imaju metaliqnost sliqnu Suncu, $0.00$} dex{\rrm , koja je raqunata iz 21 merenja radjenih za 10 zvezda.}
  
}

\end{multicols}

\newpage

\section*{APPENDIX}
\label{sec:App_Fourier}


In Tables A1, A2 and A3 we give the Fourier parameters of the investigated stars in the Milky Way, the Fourier parameters of the ACs in the LMC using the V-band data from the OGLE-III survey and the Fourier parameters of the BLHs in the LMC using the V-band data from the OGLE-III survey, respectively.


\clearpage
\thispagestyle{plain}
\begin{landscape}
\centering 
\noindent
\parbox{\linewidth}{
{\bf Table A.1.} The Fourier parameters of the investigated stars in the Milky Way.}
\vskip.25cm
\resizebox{\linewidth}{!}{
\begin{tabular}{c c c c c c c c c c c c c c c c c c c c c c c c}
\hline
\hline
Name & $f_0$ & log P[days] & $A_1$ & $A_{1err}$ & $A_2$ & $A_{2err}$ & $A_3$ & $A_{3err}$ & $R_{21}$ & $R_{21err}$ & $R_{31}$ & $R_{31err}$ & $\phi_1$ & $\phi_{1err}$ & $\phi_2$ & $\phi_{2err}$ & $\phi_3$ & $\phi_{3err}$ & $\phi_{21}$ & $\phi_{21err}$ & $\phi_{31}$ & $\phi_{31err}$ & Source\\
\hline
V742 Cyg	 & 1.06785011 & -0.0285102967 & 0.33 & 0.01 & 0.12 & 0.01 & 0.04 & 0.01 & 0.36 & 0.04 & 0.12 & 0.04 & 0.613 & 0.005 & 0.77 & 0.02 & 0.78 & 0.16 & 4.97 & 0.02 & 2.74 & 0.16 & Schmidt and Reiswig (1993)\\
FY Aqr & 0.977620241	 & 0.0098298152	 & 0.287	 & 0.005	 & 0.118	 & 0.005	 & 0.091	 & 0.005	 & 0.41	 & 0.02	 & 0.32	 & 0.02 & 0.751 & 0.003 & 0.873 & 0.007 & 0.009 & 0.009 & 3.896 & 0.007 & 1.605 & 0.009 & Pojmanski (1997)\\
V563 Cen	 & 0.928577098 & 0.0321820318 & 0.437 & 0.007 & 0.206 & 0.008	 & 0.153	 & 0.007	 & 0.47 & 0.02 & 0.35	 & 0.02 & 0.621 & 0.003 & 0.635 & 0.006 & 0.666 & 0.009 & 4.043 & 0.007 & 1.914 & 0.009 & Drake et al. (2014a)\\
FY Vir & 0.924247177	 & 0.0342118673 & 0.404 & 0.006 & 0.174 & 0.007 & 0.154 & 0.007 & 0.43 & 0.02	 & 0.38 & 0.02	 & 0.378	 & 0.003	 & 0.155	 & 0.006	 & 0.909	 & 0.007	 & 4.076	 & 0.007	 & 1.728	 & 0.008 & Drake et al. (2014a)\\
PP Tel & 0.916880399	 & 0.0376873115 & 0.36 & 0.02 & 0.20	 & 0.02 & 0.14 & 0.02 & 0.55	 & 0.07 & 0.40 & 0.06	 & 0.841 & 0.009 & 0.10 & 0.02 & 0.30	 & 0.02 & 4.19 & 0.02 & 1.74	 & 0.03 & Pojmanski (1997)\\
BX Del & 0.915927138	& 0.0381390731 & 0.359 & 0.007 & 0.12 & 0.02 & 0.064 & 0.006 & 0.35 & 0.06 & 0.18 & 0.02 & 0.727 & 0.003 & 0.96 & 0.07 & 0.98 & 0.03 & 4.77 & 0.07 & 1.87 & 0.03 & Pojmanski (1997)\\
V716 Oph & 0.896124749 & 0.0476315282 & 0.491 & 0.008 & 0.231 & 0.005 & 0.161 & 0.004 & 0.47 & 0.01 & 0.329 & 0.009 & 0.579 & 0.002 & 0.541 & 0.002 & 0.529 & 0.004 & 3.984 & 0.003 & 1.841 & 0.005 & Pojmanski (1997)\\
DF Hyi & 0.890678094 & 0.0502792289 & 0.51 & 0.03 & 0.28 & 0.03 & 0.17 & 0.05 & 0.55 & 0.07 & 0.33 & 0.09 & 0.852 & 0.009 & 0.01 & 0.02 & 0.27 & 0.17 & 3.52 & 0.02 & 1.38 & 0.17 & Pojmanski (1997)\\
BQ CrA & 0.89886751759 & 0.0521979414 & 0.27 & 0.04 & 0.21 & 0.04 & 0.16 & 0.04 & 0.77 & 0.19 & 0.59 & 0.16 & 0.41 & 0.03 & 0.19 & 0.06 & 0.81 & 0.08 & 3.91 & 0.07 & 0.46 & 0.08 & Pojmanski (1997)\\
BH Cet & 0.878962235 & 0.0560297842 & 0.420 & 0.004 & 0.198 & 0.004 & 0.136 & 0.004 & 0.47 & 0.01 & 0.323 & 0.009 & 0.145 & 0.001 & 0.679 & 0.003 & 0.229 & 0.005 & 4.019 & 0.003 & 1.854 & 0.005 & Drake et al. (2014a)\\
BF Ser & 0.858045833 & 0.0664895134 & 0.482 & 0.006 & 0.245 & 0.006 & 0.162 & 0.006 & 0.51 & 0.01 & 0.34 & 0.01 & 0.312 & 0.002 & 0.039 & 0.004 & 0.790 & 0.006 & 4.186 & 0.004 & 2.233 & 0.006 & Pojmanski (1997)\\
BI Tel & 0.857323302 & 0.0668604377 & 0.472 & 0.007 & 0.238 & 0.007 & 0.148 & 0.007 & 0.50 & 0.02 & 0.31 & 0.02 & 0.776 & 0.002 & 0.950 & 0.005 & 0.139 & 0.008 & 4.066 & 0.005 & 1.947 & 0.008 & Pojmanski (1997)\\
V2733 Oph & 0.853554943 & 0.0687685183 & 0.60 & 0.04 & 0.25 & 0.05 & 0.09 & 0.03 & 0.42 & 0.09 & 0.15 & 0.05 & 0.78 & 0.01 & 0.99 & 0.06 & 0.15 & 0.24 & 4.28 & 0.06 & 1.98 & 0.24 & Pojmanski (1997)\\
CE Her & 0.826825468 & 0.0825861546 & 0.37 & 0.01 & 0.18 & 0.01 & 0.14 & 0.01 & 0.48 & 0.04 & 0.38 & 	0.04 & 0.875 & 0.006 & 	0.49 & 0.01 & 0.48 & 0.01 & 6.22 & 0.01 & 2.26 & 0.01 & Pojmanski (1997)\\
BV Cha & 0.807728303 & 0.0927346991 & 0.380 & 0.002 & 0.139 & 0.003 & 0.106 & 0.002 & 0.365 & 0.008 & 	0.279 & 0.007 & 0.851 & 0.001 & 0.205 & 0.003 & 0.468 & 0.005 & 4.731 & 0.003 & 2.608 & 0.005 & Pojmanski (1997)\\
VY Pyx & 0.806485292 & 0.0934035485 & 0.118 & 0.001 & 0.019 & 0.001 & 0.005 & 0.001 & 0.16 & 0.01 & 0.04 & 0.01 & 0.655 & 0.002 & 0.85 & 0.01 & 0.45 & 0.22 & 4.97  & 0.01 & 6.16 & 0.22 & Pojmanski (1997)\\
V527 Sgr & 0.797035422 & 0.0985223772 & 0.36 & 0.01 & 0.12 & 0.02 & 0.03 & 0.01 & 0.32 & 0.05 & 0.08 & 0.04 & 0.949 & 0.006 & 0.32 & 0.02 & 0.42 & 0.04 & 4.25 & 0.02 & 0.46 & 0.04 & Kwee and Diethelm (1984)\\
BL Her & 0.764852872 & 0.1164220982 & 0.342 & 0.002 & 0.121 & 0.002 & 0.082 & 0.002 & 0.353 & 0.006 & 0.241 & 0.007 & 0.798 & 	0.001 & 0.101 & 0.003 & 0.408 & 0.004 & 4.740 & 0.003 & 3.223 & 0.004 & Pojmanski (1997)\\
VX Cap & 0.750502164 & 0.1246480512 & 0.479 & 0.005 & 0.227 & 0.005 & 0.130 & 0.005 & 0.47 & 0.01 & 	0.27 & 0.01 & 0.222 & 0.002 & 0.891 & 0.003 & 0.531 & 0.006 & 4.446 & 0.004 & 2.393 & 0.007 & Drake et al. (2014a)\\
V5614 Sgr & 0.738490213 & 0.1316552559 & 0.31 & 0.01 & 0.12 & 0.01 & 0.07 & 0.01 & 0.37 & 0.04 & 0.22 & 	0.04 & 0.898 & 	0.005 & 0.34 & 0.01 & 0.79 & 0.02 & 4.96 & 0.02 & 3.74 & 0.02 & Soszy\'{n}ski et al. (2011b)\\
XX Vir & 0.741726923 & 0.1297559568 & 0.392 & 0.006 & 0.193 & 0.006 & 0.132 & 0.006 & 0.49 & 0.02 & 0.34 & 	0.02 & 0.465 & 0.002 & 0.343 & 0.005 & 0.248 & 0.007 & 4.163 & 0.005 & 2.215 & 0.007 & Pojmanski (1997)\\
V1149 Her & 	0.709424427 & 0.140938619 & 	0.336 & 0.004 & 0.173 & 0.004 & 0.123 & 0.004 & 0.514 & 0.0126 & 0.37 & 0.01 & 0.481 & 0.002 & 0.392 & 0.003 & 0.317 & 0.005 & 4.274 & 0.004 & 	2.356 & 0.005 & Drake et al. (2014a)\\
HQ CrA & 0.706485722 & 0.1508966108 & 0.39 & 0.02 & 0.12 & 0.01 & 0.01 & 0.01 & 0.31 & 0.04 & 0.03 & 0.03 & 0.040 & 0.006 & 0.43 & 0.02 & 0.69 & 0.19 & 3.79 & 0.02 & 0.43 & 0.19 & Kwee and Diethelm (1984)\\
MQ Aql & 0.675328868 & 0.1704846853 & 0.29 & 0.02 & 0.18 & 0.02 & 0.17 & 0.02 & 0.64 & 0.10 & 0.59 & 0.09 & 	0.21 & 0.01 & 0.75 & 0.02 & 0.46 & 0.02 & 3.58 & 0.03 & 2.01 & 0.03 & Pojmanski (1997)\\
KZ Cen & 0.657879317 & 0.1818537671 & 0.384 & 0.004 & 0.144 & 0.005 & 0.053 & 0.004 & 0.38 & 0.01 & 0.14 & 0.01 & 0.311 & 0.002 & 0.991 & 0.005 & 0.46 & 0.01 & 3.881 & 0.00 & 0.15 & 0.01 & Pojmanski (1997)\\
V2022 Sgr & 0.653820108 & 0.184541727 & 0.38 & 0.01 & 0.14 & 0.08 & 0.05 & 0.08 & 0.37 & 0.21 & 0.13 & 0.20 & 0.952 & 0.006 & 0.54 & 0.01 & 0.60 & 0.14 & 5.59 & 0.02 & 1.56 & 0.14 & Kwee and Diethelm (1984)\\
SW Tau & 0.631499487 & 0.1996269979 & 0.353 & 0.004 & 0.115 & 0.004 & 0.0305 & 0.004 & 0.33 & 0.01 & 	0.09 & 0.01 & 0.026 & 0.002 & 0.515 & 0.005 & 0.15 & 0.02 & 4.474 & 0.005 & 3.584 & 0.020 & Pojmanski (1997)\\
V745 Oph & 0.626497323 & 0.2030807804 & 0.25 & 0.02 & 0.09 & 0.03 & 0.05 & 0.02 & 0.37 & 0.12 & 0.18 & 0.07 & 0.42 & 0.01 & 0.27 & 0.25 & 0.34 & 0.07 & 4.18 & 0.25 & 3.58 & 0.07& Pojmanski (1997)\\
NW Lyr & 0.624493109 & 0.2044723495 & 0.40 & 0.03 & 0.16 & 0.03 & 0.08 & 0.03 & 0.40 & 0.07 & 0.20 & 0.07 & 0.890 & 0.010 & 0.29 & 0.02 & 0.42 & 0.06 & 4.75 & 0.03 & 1.54 & 0.06 & Schmidt et al. (2005)\\
V971 Aql & 0.615557473 & 0.2107313919 & 0.363 & 0.005 & 0.091 & 0.005 & 0.07 & 0.005 & 0.25 & 0.01 & 0.19 & 0.01 & 0.657 & 0.002 & 0.827 & 0.009 & 0.71 & 0.01 & 4.791 & 0.009 & 	1.52 & 0.01 & Pojmanski (1997)\\
VZ Aql & 0.593633762 & 0.2264814072 & 0.6 & 0.09 & 0.42 & 0.05 & 0.20 & 0.06 & 0.65 & 0.12 & 	0.30 & 	0.09 & 0.09 & 0.04 & 0.91 & 0.07 & 0.31 & 0.16 & 6.15 & 0.07 & 3.40 & 0.16 & INTEGRAL\\
V1437 Sgr & 0.572075542 & 0.2425466193 & 0.41 & 0.01 & 0.16 & 0.01 & 0.09 & 0.01 & 0.38 & 0.03 & 0.21 & 0.03 & 0.983 & 0.005 & 	0.41 & 0.01 & 0.69 & 0.02 & 4.33 & 0.02 & 1.51 & 0.03 & Soszy\'{n}ski et al. (2011b)\\
V714 Cyg & 0.529782964 & 0.2759020112 & 0.42 & 0.01 & 0.15 & 0.01 & 0.11 & 0.01 & 0.36 & 0.03 & 0.27 & 0.03 & 0.383 & 0.004 & 0.18 & 0.01 & 0.94 & 0.02 & 4.19 & 0.01 & 1.83 & 0.02 & Schmidt et al. (2005)\\
V439 Oph & 0.52826286 & 0.2771499217 & 0.30 & 0.02 & 0.05 & 0.02 & 0.04 & 0.02 & 0.16 & 0.07 & 0.15 & 0.08 & 0.47 & 0.06 & 0.38 & 0.15 & 0.84 & 0.34 & 4.34 & 0.16 & 5.78 & 0.35 & Schmidt et al. (2005)\\
RT TrA & 0.513842147 & 0.2891702764 & 0.323 & 0.001 & 0.034 & 0.001 & 0.031 & 0.001 & 0.105 & 0.004 & 0.095 & 0.004 & 0.9926 & 0.0006 & 0.503 & 0.006 & 0.693 & 0.006 & 4.821 & 0.005 & 1.350 & 0.007 & Pojmanski (1997)\\
GK Cen & 0.512882982 & 0.289981711 & 0.267 & 0.007 & 0.049 & 0.006 & 0.031 & 0.007 & 0.18 & 0.02 & 	0.12 & 0.02 & 0.796 & 0.004 & 0.09 & 0.02 & 0.23 & 0.03 & 4.72 & 0.02 & 2.189 & 0.04& Pojmanski (1997)\\
AT Tel & 0.507627684 & 0.2944547012 & 0.285 & 0.03 & 0.12 & 0.02 & 0.06 & 0.02 & 0.43 & 0.09 & 0.19 & 0.07 & 0.34 & 0.01 & 0.94 & 0.03 & 0.50 & 0.24 & 3.19 & 0.04 & 6.13 & 0.24 & Pojmanski (1997)\\
V477 Oph & 	0.496086274 & 0.3044427891 & 0.34 & 0.02 & 0.12 & 0.04 & 0.06 & 0.02 & 0.37 & 0.13 & 0.19 & 	0.06 & 0.93 & 0.01 & 0.33 & 0.20 & 0.23 & 0.10 & 4.50 & 0.20 & 5.84 & 0.10 & Pojmanski (1997)\\
V1287 Sco & 0.491265871 & 0.3086834059 & 0.22 & 0.06 & 0.14 & 0.05 & 0.08 & 0.03 & 0.63 & 0.27 & 0.39 & 	0.16 & 0.51 & 0.13 & 0.97 & 0.12 & 0.04 & 0.07 & 1.21 & 0.18 & 0.01 & 0.15 & Pojmanski (1997)\\
V553 Cen & 0.485305739 & 0.3139845729 & 0.230 & 0.002 & 0.012 & 0.003 & 0.020 & 0.003 & 0.05 & 0.01 & 	0.09 & 0.01 & 0.284 & 0.002 & 0.005 & 0.177 & 0.57 & 0.06 & 4.31 & 0.17 & 1.37 & 0.06 & Pojmanski (1997)\\
V5608 Sgr & 0.45193215 & 0.3449267623 & 0.38 & 0.01 & 0.066 & 0.009 & 0.0376 & 0.009 & 0.17 & 0.03 & 	0.10 & 0.03 & 0.842 & 0.004 & 0.22 & 0.03 & 0.20 & 0.05 & 4.93 & 0.03 & 4.23 & 0.05& Soszy\'{n}ski et al. (2011b)\\
UY Eri & 0.451814735	 & 0.3450396096 & 0.269 & 0.002 & 0.111 & 0.002 & 0.038 & 0.002 & 0.415 & 0.009 & 0.143 & 0.007 & 0.952 & 0.001 & 0.369 & 0.003 & 0.789 & 0.009 & 4.494 & 	0.004 & 2.723 & 0.009 & Pojmanski (1997)\\
UX Nor & 0.419109544 & 0.3776724493 & 0.38 & 0.02 & 0.21 & 0.02 & 0.13 & 0.02 & 0.56 & 0.07 & 0.35 & 0.06 & 0.62 & 0.01 & 0.69 & 0.02 & 0.86 & 0.03 & 4.45 & 0.02 & 3.25 & 0.03 & Pojmanski (1997)\\
V617 Ara & 0.396512214 & 0.4017434303 & 0.231 & 0.003 & 0.061 & 0.003 & 0.028 & 0.003 & 0.27 & 0.01 & 0.12 & 0.01 & 0.996 & 0.002 & 0.446 & 0.007 & 0.91 & 0.01 & 4.426 & 0.007 & 2.65 & 0.01 & Pojmanski (1997)\\
V351 Cep & 0.356303349 & 0.4481800956 & 0.149 & 0.007 & 0.022 & 0.007 & 0.011 & 0.006 & 0.15 & 0.05 & 0.08 & 	0.04 & 0.959 & 0.008 & 0.12 & 0.05 & 0.08 & 0.12 & 2.85 & 0.05 & 4.42 & 0.12 & Szabados (1977) Henden (1980)\\
V465 Oph & 0.351651936 & 0.4538869873 & 0.40 & 0.01 & 0.14 & 0.01 & 0.07 & 0.01 & 0.34 & 0.03 & 0.17 & 0.03 & 0.673 & 0.004 & 	0.83 & 0.01 & 0.97 & 0.03 & 4.65 & 0.01 & 2.83 & 0.03 & Pojmanski (1997)\\
DQ And & 0.312444362 & 0.5052273077 & 0.316 & 0.005 & 0.133 & 0.005 & 0.078 & 0.006 & 0.42 & 0.02 & 0.25 & 0.02 & 0.505 & 0.003 & 0.433 & 0.007 & 0.37 & 0.01 & 4.228 & 0.007 & 2.22 & 0.02 & Schmidt et al. (2005)\\
BE CrA & 0.299683177 & 0.5233376359 & 0.47 & 0.03 & 0.18 & 0.06 & 0.10 & 0.03 & 0.38 & 0.14 & 0.21 & 	0.06 & 0.584 & 0.006 & 0.67 & 0.21 & 0.85 & 0.09 & 4.76 & 0.21 & 3.73 & 0.09 & Pojmanski (1997)\\
FM Del & 0.299679663 & 0.5233427283 & 0.47 & 0.03 & 0.18 & 0.06 & 0.10 & 0.03 & 0.37 & 0.14 & 0.21 & 	0.06 & 0.597 & 0.006 & 0.65 & 0.20 & 0.79 & 0.08 & 4.45 & 0.20 & 3.16 & 0.08 & Pojmanski (1997)\\
BD Cas & 0.273901489 & 0.5624056068 & 0.160 & 0.002 & 0.016 & 0.002 & 0.007 & 0.002 & 0.10 & 0.01 & 0.05 & 0.01 & 0.080 & 0.002 & 0.54 & 0.03 & 0.74 & 0.06 & 3.98 & 0.03 & 0.02 & 0.06 & INTEGRAL\\
V5609 Sgr & 0.282290865 & 0.5493031755 & 0.302 & 0.010 & 0.112 & 	0.010 & 0.056 & 0.009 & 0.37 & 0.03 & 	0.19 & 0.03 & 0.714 & 0.005 & 0.01 & 0.01 & 0.27 & 0.03 & 5.25 & 0.01 & 3.96 & 0.03 & Soszy\'{n}ski et al. (2011b)\\
QY Cyg & 0.256861546 & 0.5903009078 & 0.19 & 0.03 & 0.14 & 0.03 & 0.15 & 0.03 & 0.70 & 0.20 & 0.78 & 	0.20 & 0.30 & 0.02 & 0.88 & 0.03 & 0.74 & 0.03 & 3.40 & 0.04 & 2.22 & 0.04 & INTEGRAL\\
V383 Cyg & 0.216813043 & 0.6639145951 & 0.248 & 0.006 & 0.076 & 0.006 & 0.039 & 0.007 & 0.31 & 0.03 & 0.16 & 0.03 & 0.646 & 0.004 & 0.71 & 0.01 & 0.81 & 0.03 & 4.21 & 0.01 & 2.33 & 0.03 & Berdnikov (2008), AAVSO\\
V675 Cen & 0.216011549 & 0.6655230288 & 0.298 & 0.009 & 0.114 & 	0.006 & 0.05 & 0.02 & 0.38 & 0.02 & 0.17 & 0.07 & 0.317 & 0.003 & 0.57 & 0.01 & 0.18 & 0.22 & 1.15 & 0.01 & 4.60 & 0.22 & Pojmanski (1997)\\
V394 Cep & 0.175807826 & 0.7549617964 & 0.371 & 0.006 & 0.15 & 0.03 & 0.060 & 0.008 & 0.39 & 0.07 & 	0.16 & 0.02 & 0.065 & 0.004 & 0.59 & 0.19 & 0.07 & 0.02 & 4.45 & 0.19 & 2.33 & 0.02 & INTEGRAL\\
AB Ara & 0.167854242 & 0.7750676788 & 0.34 & 0.01 & 	0.13 & 0.01 & 0.04 & 0.01 & 0.39 & 0.03 & 0.13 & 	0.03 & 0.639 & 	0.005 & 0.79 & 0.01 & 0.97 & 0.06 & 4.82 & 0.01 & 3.46 & 0.06 & Pojmanski (1997)\\
TX Del & 0.162177362 & 0.7900097682 & 0.294 & 0.002 & 0.072 & 0.002 & 0.022 & 0.002 & 0.245 & 0.009 & 0.076 & 0.008 & 0.776 & 0.001 & 0.091 & 0.006 & 0.36 & 0.02 & 4.960 & 0.006 & 3.32 & 0.19 & Pojmanski (1997)\\
UY CrA & 0.142949436 & 0.8448175538 & 0.52 & 0.02 & 0.13 & 0.02 & 0.14 & 0.02 & 0.25 & 0.05 & 0.27 & 0.04 & 	0.019 & 0.007 & 0.43 & 0.03 & 0.05 & 0.03 & 4.00 & 0.03 & 3.10 & 	0.033 & INTEGRAL\\
IT Cep & 0.136081047 & 0.8662023579 & 0.232 & 0.005 & 0.077 & 0.004 & 0.016 & 0.004 & 0.33 & 0.02 & 0.07 & 0.02 & 0.587 & 0.003 & 0.70 & 0.01 & 0.73 & 0.05 & 4.87 & 0.01 & 	2.96 & 0.05 & Schmidt et al. (2005)\\
\hline
\end{tabular}
}
\end{landscape}
\clearpage



\clearpage
\thispagestyle{plain}
\begin{landscape}
\centering 
\noindent
\parbox{\linewidth}{
{\bf Table A.2.} The Fourier parameters of the ACs in the LMC using the V-band data from the OGLE-III survey. }
\vskip.25cm
\resizebox{\linewidth}{!}{
\begin{tabular}{p{1cm} c c c c c c c c c c c c c c c c c c c c c c}
\hline
\hline
OGLE-LMC-ACEP & $f_0$ & log P[days] & $A_1$ & $A_{1err}$ & $A_2$ & $A_{2err}$ & $A_3$ & $A_{3err}$ & $R_{21}$ & $R_{21err}$ & $R_{31}$ & $R_{31err}$ & $\phi_1$ & $\phi_{1err}$ & $\phi_2$ & $\phi_{2err}$ & $\phi_3$ & $\phi_{3err}$ & $\phi_{21}$ & $\phi_{21err}$ & $\phi_{31}$ & $\phi_{31err}$\\
\hline
3 & 2.61930268 & -0.4181856873 & 0.379 & 0.015 & 0.128 & 0.031 & 0.052 & 0.018 & 0.338 & 0.082 & 0.136 & 0.046 & 0.705 & 0.005 & 0.704 & 0.151 & 0.967 & 0.232 & 3.417 & 0.151 & 2.214 & 0.232\\
6 & 1.17707809 & -0.0708052759 & 0.256 & 0.011 & 0.12 & 0.009 & 0.053 & 0.017 & 0.469 & 0.038 & 0.204 & 0.065 & 0.564 & 0.004 & 0.647 & 0.011 & 0.782 & 0.287 & 4.834 & 0.011 & 3.713 & 0.287\\
7 & 1.11555371 & -0.0474904849 & 0.44 & 0.03 & 0.165 & 0.026 & 0.099 & 0.033 & 0.376 & 0.063 & 0.224 & 0.076 & 0.183 & 0.008 & 0.721 & 0.027 & 0.462 & 0.253 & 3.803 & 0.028 & 2.602 & 0.253\\
8 & 1.33497031 & -0.125471607 & 0.2 & 0.007 & 0.062 & 0.006 & 0.021 & 0.004 & 0.307 & 0.028 & 0.103 & 0.021 & 0.04 & 0.005 & 0.29 & 0.064 & 0.887 & 0.05 & 2.888 & 0.064 & 1.678 & 0.05\\
14 & 0.436427013 & 0.3600883762 & 0.348 & 0.01 & 0.149 & 0.038 & 0.063 & 0.01 & 0.426 & 0.109 & 0.18 & 0.029 & 0.916 & 0.011 & 0.328 & 0.232 & 0.751 & 0.027 & 4.691 & 0.232 & 3.167 & 0.029\\
16 & 0.646943545 & 0.189133616 & 0.421 & 0.009 & 0.211 & 0.013 & 0.143 & 0.009 & 0.501 & 0.033 & 0.339 & 0.023 & 0.082 & 0.003 & 0.423 & 0.01 & 0.821 & 0.01 & 3.2 & 0.01 & 0.475 & 0.011\\
17 & 1.07527992 & -0.0315215358 & 0.335 & 0.015 & 0.135 & 0.017 & 0.089 & 0.015 & 0.403 & 0.052 & 0.264 & 0.044 & 0.492 & 0.007 & 0.353 & 0.084 & 0.411 & 0.026 & 3.883 & 0.085 & 2.729 & 0.027\\
19 & 1.09962029 & -0.0412427448 & 0.449 & 0.007 & 0.192 & 0.007 & 0.125 & 0.006 & 0.427 & 0.016 & 0.279 & 0.014 & 0.421 & 0.002 & 0.184 & 0.005 & 0.973 & 0.008 & 3.723 & 0.006 & 1.333 & 0.009\\
20 & 2.6184651 & -0.4180467898 & 0.345 & 0.009 & 0.126 & 0.039 & 0.047 & 0.008 & 0.366 & 0.114 & 0.135 & 0.023 & 0.687 & 0.004 & 0.626 & 0.224 & 0.565 & 0.08 & 3.155 & 0.224 & 0.031 & 0.08\\
21 & 0.771695927 & 0.1125537919 & 0.381 & 0.004 & 0.181 & 0.004 & 0.15 & 0.003 & 0.474 & 0.011 & 0.393 & 0.009 & 0.671 & 0.002 & 0.727 & 0.005 & 0.807 & 0.003 & 3.992 & 0.005 & 1.859 & 0.004\\
22 & 1.56018852 & -0.193177078 & 0.296 & 0.018 & 0.132 & 0.017 & 0.095 & 0.011 & 0.446 & 0.061 & 0.322 & 0.041 & 0.192 & 0.008 & 0.664 & 0.017 & 0.486 & 0.064 & 3.33 & 0.019 & 2.577 & 0.065\\
23 & 1.38229242 & -0.1405999265 & 0.25 & 0.003 & 0.122 & 0.004 & 0.057 & 0.003 & 0.486 & 0.014 & 0.226 & 0.012 & 0.744 & 0.002 & 0.911 & 0.004 & 0.987 & 0.009 & 4.229 & 0.004 & 1.614 & 0.009\\
24 & 1.25871428 & -0.0999271593 & 0.288 & 0.013 & 0.13 & 0.011 & 0.092 & 0.012 & 0.45 & 0.042 & 0.319 & 0.043 & 0.867 & 0.007 & 0.011 & 0.015 & 0.351 & 0.022 & 3.314 & 0.016 & 1.58 & 0.022\\
26 & 0.575126634 & 0.2402365198 & 0.339 & 0.005 & 0.167 & 0.005 & 0.108 & 0.005 & 0.492 & 0.015 & 0.319 & 0.015 & 0.666 & 0.002 & 0.775 & 0.005 & 0.858 & 0.007 & 4.354 & 0.006 & 2.264 & 0.007\\
32 & 0.759864208 & 0.1192640116 & 0.248 & 0.005 & 0.112 & 0.027 & 0.069 & 0.005 & 0.451 & 0.11 & 0.278 & 0.02 & 0.549 & 0.005 & 0.479 & 0.15 & 0.472 & 0.016 & 3.97 & 0.15 & 2.05 & 0.016\\
35 & 2.24191287 & -0.3506187301 & 0.339 & 0.007 & 0.128 & 0.034 & 0.038 & 0.007 & 0.376 & 0.099 & 0.112 & 0.019 & 0.055 & 0.004 & 0.511 & 0.219 & 0.008 & 0.082 & 4.09 & 0.219 & 2.157 & 0.083\\
36 & 0.794929717 & 0.0996712674 & 0.178 & 0.013 & 0.061 & 0.012 & 0.031 & 0.014 & 0.341 & 0.07 & 0.173 & 0.075 & 0.438 & 0.007 & 0.271 & 0.023 & 0.146 & 0.047 & 4.058 & 0.024 & 2.092 & 0.047\\
38 & 0.748949532 & 0.1255474463 & 0.285 & 0.013 & 0.134 & 0.039 & 0.094 & 0.014 & 0.469 & 0.138 & 0.329 & 0.052 & 0.862 & 0.01 & 0.138 & 0.25 & 0.376 & 0.085 & 4.173 & 0.25 & 1.827 & 0.086\\
39 & 1.00765091 & -0.0033101014 & 0.331 & 0.011 & 0.143 & 0.034 & 0.108 & 0.015 & 0.43 & 0.102 & 0.325 & 0.044 & 0.898 & 0.007 & 0.221 & 0.229 & 0.54 & 0.025 & 4.248 & 0.229 & 2.189 & 0.026\\
40 & 1.04104059 & -0.0174676629 & 0.3 & 0.008 & 0.117 & 0.008 & 0.084 & 0.007 & 0.389 & 0.027 & 0.278 & 0.025 & 0.206 & 0.004 & 0.8 & 0.01 & 0.421 & 0.015 & 4.012 & 0.011 & 1.913 & 0.015\\
41 & 1.13876929 & -0.0564357467 & 0.308 & 0.007 & 0.121 & 0.008 & 0.083 & 0.019 & 0.392 & 0.025 & 0.269 & 0.062 & 0.159 & 0.004 & 0.705 & 0.008 & 0.243 & 0.276 & 4.009 & 0.009 & 1.68 & 0.276\\
44 & 0.764227056 & 0.116777591 & 0.45 & 0.005 & 0.237 & 0.005 & 0.16 & 0.005 & 0.527 & 0.012 & 0.356 & 0.011 & 0.814 & 0.002 & 0.055 & 0.003 & 0.293 & 0.005 & 4.249 & 0.004 & 2.209 & 0.005\\
46 & 0.791317111 & 0.1016494432 & 0.328 & 0.004 & 0.097 & 0.004 & 0.095 & 0.004 & 0.296 & 0.012 & 0.289 & 0.011 & 0.054 & 0.002 & 0.284 & 0.004 & 0.045 & 0.006 & 2.676 & 0.004 & 2.405 & 0.006\\
48 & 0.646873474 & 0.1891806574 & 0.398 & 0.004 & 0.21 & 0.004 & 0.136 & 0.004 & 0.528 & 0.01 & 0.34 & 0.009 & 0.63 & 0.002 & 0.681 & 0.003 & 0.74 & 0.004 & 4.223 & 0.003 & 2.211 & 0.004\\
49 & 1.55087039 & -0.1905755043 & 0.226 & 0.01 & 0.108 & 0.009 & 0.08 & 0.01 & 0.477 & 0.045 & 0.354 & 0.045 & 0.399 & 0.007 & 0.2 & 0.015 & 0.935 & 0.021 & 4.103 & 0.017 & 1.504 & 0.022\\
51 & 1.41125232 & -0.1496046689 & 0.226 & 0.008 & 0.055 & 0.008 & 0.051 & 0.011 & 0.242 & 0.036 & 0.224 & 0.049 & 0.068 & 0.007 & 0.289 & 0.041 & 0.051 & 0.109 & 2.54 & 0.042 & 2.193 & 0.109\\
52 & 0.792038365 & 0.1012537814 & 0.413 & 0.029 & 0.202 & 0.024 & 0.171 & 0.047 & 0.488 & 0.067 & 0.413 & 0.117 & 0.232 & 0.01 & 0.791 & 0.029 & 0.479 & 0.192 & 3.621 & 0.03 & 1.78 & 0.193\\
53 & 0.529628975 & 0.2760282635 & 0.437 & 0.01 & 0.205 & 0.008 & 0.108 & 0.006 & 0.468 & 0.02 & 0.247 & 0.015 & 0.104 & 0.005 & 0.67 & 0.011 & 0.149 & 0.02 & 4.475 & 0.012 & 2.124 & 0.02\\
57 & 0.584789332 & 0.2330005586 & 0.403 & 0.004 & 0.236 & 0.004 & 0.171 & 0.003 & 0.585 & 0.01 & 0.424 & 0.007 & 0.156 & 0.001 & 0.815 & 0.002 & 0.457 & 0.003 & 4.726 & 0.002 & 3.072 & 0.003\\
60 & 0.783862684 & 0.1057600098 & 0.475 & 0.006 & 0.245 & 0.006 & 0.159 & 0.006 & 0.516 & 0.013 & 0.335 & 0.012 & 0.201 & 0.002 & 0.819 & 0.004 & 0.441 & 0.006 & 4.188 & 0.004 & 2.125 & 0.006\\
62 & 0.94421245 & 0.0249302774 & 0.485 & 0.008 & 0.208 & 0.007 & 0.129 & 0.008 & 0.429 & 0.016 & 0.266 & 0.017 & 0.414 & 0.002 & 0.214 & 0.007 & 0.016 & 0.009 & 3.996 & 0.007 & 1.727 & 0.009\\
63 & 1.11978242 & -0.049133645 & 0.331 & 0.016 & 0.115 & 0.014 & 0.055 & 0.01 & 0.348 & 0.044 & 0.166 & 0.03 & 0.208 & 0.005 & 0.811 & 0.016 & 0.716 & 0.036 & 4.051 & 0.017 & 3.717 & 0.036\\
65 & 0.756691193 & 0.1210813207 & 0.468 & 0.006 & 0.234 & 0.006 & 0.145 & 0.006 & 0.501 & 0.014 & 0.309 & 0.012 & 0.425 & 0.002 & 0.283 & 0.004 & 0.137 & 0.006 & 4.294 & 0.005 & 2.284 & 0.006\\
\hline
\end{tabular}
}
\end{landscape}
\clearpage


\clearpage
\thispagestyle{plain}
\begin{landscape}
\centering 
\noindent
\parbox{\linewidth}{
{\bf Table A.3.} The Fourier parameters of the BLHs in the LMC using the V-band data from the OGLE-III survey. }
\vskip.25cm
\resizebox{\linewidth}{!}{
\begin{tabular}{p{1cm} c c c c c c c c c c c c c c c c c c c c c c}
\hline
\hline
OGLE-LMC-T2CEP & $f_0$ & log P[days] & $A_1$ & $A_{1err}$ & $A_2$ & $A_{2err}$ & $A_3$ & $A_{3err}$ & $R_{21}$ & $R_{21err}$ & $R_{31}$ & $R_{31err}$ & $\phi_1$ & $\phi_{1err}$ & $\phi_2$ & $\phi_{2err}$ & $\phi_3$ & $\phi_{3err}$ & $\phi_{21}$ & $\phi_{21err}$ & $\phi_{31}$ & $\phi_{31err}$\\
\hline
7 & 0.804721291 & 0.0943545081 & 0.363 & 0.008 & 0.098 & 0.01 & 0.045 & 0.007 & 0.27 & 0.028 & 0.123 & 0.018 & 0.246 & 0.003 & 0.889 & 0.012 & 0.871 & 0.035 & 4.07 & 0.013 & 3.984 & 0.035\\
9 & 0.567734944 & 0.2458543742 & 0.413 & 0.01 & 0.169 & 0.011 & 0.087 & 0.011 & 0.408 & 0.028 & 0.209 & 0.025 & 0.738 & 0.005 & 0.884 & 0.011 & 0.886 & 0.018 & 4.142 & 0.012 & 1.098 & 0.019\\
24 & 0.80212731 & 0.095756697 & 0.37 & 0.014 & 0.096 & 0.013 & 0.033 & 0.014 & 0.258 & 0.036 & 0.089 & 0.038 & 0.023 & 0.006 & 0.448 & 0.027 & 0.207 & 0.275 & 4.099 & 0.028 & 4.022 & 0.275\\
30 & 0.254110433 & 0.5949775038 & 0.256 & 0.016 & 0.082 & 0.028 & 0.04 & 0.016 & 0.321 & 0.11 & 0.155 & 0.062 & 0.269 & 0.007 & 0.142 & 0.395 & 0.97 & 0.226 & 5.372 & 0.395 & 4.174 & 0.226\\
48 & 0.691842499 & 0.1599927633 & 0.426 & 0.026 & 0.147 & 0.046 & 0.064 & 0.033 & 0.345 & 0.11 & 0.15 & 0.076 & 0.968 & 0.003 & 0.318 & 0.293 & 0.422 & 0.378 & 3.968 & 0.293 & 0.114 & 0.378\\
49 & 0.309093736 & 0.5099097961 & 0.176 & 0.009 & 0.048 & 0.012 & 0.011 & 0.007 & 0.272 & 0.065 & 0.061 & 0.037 & 0.931 & 0.008 & 0.392 & 0.076 & 0.891 & 0.29 & 4.906 & 0.076 & 3.766 & 0.29\\
53 & 0.95877363 & 0.0182839193 & 0.447 & 0.015 & 0.203 & 0.01 & 0.088 & 0.027 & 0.454 & 0.027 & 0.196 & 0.06 & 0.793 & 0.009 & 0.076 & 0.018 & 0.345 & 0.187 & 4.649 & 0.02 & 2.936 & 0.187\\
60 & 0.808680959 & 0.0922227826 & 0.433 & 0.008 & 0.106 & 0.012 & 0.025 & 0.007 & 0.244 & 0.027 & 0.057 & 0.017 & 0.585 & 0.005 & 0.57 & 0.011 & 0.865 & 0.345 & 4.08 & 0.012 & 3.834 & 0.345\\
61 & 0.846373695 & 0.0724378428 & 0.316 & 0.005 & 0.063 & 0.006 & 0.027 & 0.005 & 0.199 & 0.017 & 0.083 & 0.015 & 0.715 & 0.003 & 0.88 & 0.013 & 0.138 & 0.024 & 4.403 & 0.013 & 3.112 & 0.025\\
68 & 0.621385688 & 0.2066387538 & 0.218 & 0.005 & 0.053 & 0.005 & 0.024 & 0.005 & 0.243 & 0.022 & 0.111 & 0.02 & 0.029 & 0.004 & 0.508 & 0.015 & 0.582 & 0.032 & 4.398 & 0.016 & 6.255 & 0.033\\
69 & 0.979185356 & 0.0091350901 & 0.179 & 0.017 & 0.051 & 0.013 & 0.045 & 0.016 & 0.282 & 0.074 & 0.249 & 0.091 & 0.015 & 0.011 & 0.59 & 0.1 & 0.172 & 0.28 & 5.089 & 0.101 & 3.947 & 0.281\\
71 & 0.867929397 & 0.0615156017 & 0.431 & 0.006 & 0.139 & 0.006 & 0.027 & 0.007 & 0.322 & 0.013 & 0.061 & 0.015 & 0.659 & 0.002 & 0.77 & 0.007 & 0.829 & 0.111 & 4.412 & 0.007 & 2.217 & 0.111\\
73 & 0.323840472 & 0.4896688762 & 0.384 & 0.019 & 0.131 & 0.014 & 0.067 & 0.013 & 0.34 & 0.039 & 0.175 & 0.035 & 0.488 & 0.005 & 0.535 & 0.161 & 0.558 & 0.02 & 5.077 & 0.162 & 3.728 & 0.02\\
76 & 0.475216778 & 0.3231082346 & 0.427 & 0.02 & 0.138 & 0.023 & 0.085 & 0.029 & 0.321 & 0.055 & 0.199 & 0.068 & 0.771 & 0.009 & 0.059 & 0.061 & 0.197 & 0.197 & 4.823 & 0.062 & 2.421 & 0.197\\
77 & 0.82386308 & 0.0841449589 & 0.151 & 0.005 & 0.039 & 0.005 & 0.011 & 0.004 & 0.255 & 0.032 & 0.07 & 0.027 & 0.183 & 0.005 & 0.863 & 0.049 & 0.446 & 0.263 & 4.692 & 0.049 & 2.497 & 0.263\\
85 & 0.293677425 & 0.5321294364 & 0.298 & 0.006 & 0.09 & 0.006 & 0.033 & 0.005 & 0.301 & 0.021 & 0.11 & 0.016 & 0.257 & 0.003 & 0.097 & 0.008 & 0.913 & 0.031 & 5.244 & 0.008 & 4.046 & 0.032\\
89 & 0.85667704 & 0.0671828726 & 0.373 & 0.02 & 0.077 & 0.022 & 0.029 & 0.009 & 0.207 & 0.06 & 0.077 & 0.024 & 0.327 & 0.058 & 0.051 & 0.24 & 0.91 & 0.059 & 4.075 & 0.247 & 2.711 & 0.082\\
90 & 0.676157034 & 0.1699524297 & 0.224 & 0.006 & 0.038 & 0.007 & 0.033 & 0.007 & 0.168 & 0.029 & 0.144 & 0.028 & 0.746 & 0.004 & 0.03 & 0.08 & 0.75 & 0.052 & 4.948 & 0.081 & 0.072 & 0.052\\
92 & 0.382183436 & 0.4177281394 & 0.347 & 0.02 & 0.088 & 0.029 & 0.039 & 0.014 & 0.253 & 0.084 & 0.112 & 0.04 & 0.44 & 0.006 & 0.425 & 0.252 & 0.372 & 0.248 & 4.989 & 0.252 & 3.467 & 0.248\\
102 & 0.789875542 & 0.1024413336 & 0.436 & 0.006 & 0.123 & 0.005 & 0.044 & 0.005 & 0.283 & 0.012 & 0.101 & 0.011 & 0.214 & 0.002 & 0.858 & 0.007 & 0.545 & 0.019 & 4.275 & 0.007 & 2.541 & 0.02\\
105 & 0.67145576 & 0.1729825963 & 0.408 & 0.005 & 0.14 & 0.006 & 0.055 & 0.006 & 0.343 & 0.013 & 0.135 & 0.014 & 0.889 & 0.002 & 0.157 & 0.006 & 0.193 & 0.035 & 3.951 & 0.006 & 0.163 & 0.035\\
107 & 0.827030392 & 0.0824785305 & 0.239 & 0.008 & 0.054 & 0.015 & 0.022 & 0.008 & 0.225 & 0.062 & 0.091 & 0.031 & 0.997 & 0.005 & 0.392 & 0.395 & 0.881 & 0.122 & 4.073 & 0.395 & 2.455 & 0.123\\
116 & 0.508472942 & 0.2937321528 & 0.284 & 0.006 & 0.038 & 0.006 & 0.036 & 0.007 & 0.134 & 0.021 & 0.125 & 0.023 & 0.205 & 0.004 & 0.922 & 0.024 & 0.468 & 0.028 & 4.786 & 0.024 & 2.219 & 0.028\\
138 & 0.717571438 & 0.1441348561 & 0.312 & 0.007 & 0.06 & 0.009 & 0.028 & 0.009 & 0.19 & 0.029 & 0.089 & 0.026 & 0.361 & 0.004 & 0.146 & 0.067 & 0.489 & 0.051 & 4.237 & 0.067 & 5.698 & 0.051\\
145 & 0.29962254 & 0.5234255186 & 0.13 & 0.004 & 0.024 & 0.004 & 0.006 & 0.004 & 0.179 & 0.03 & 0.041 & 0.025 & 0.365 & 0.004 & 0.214 & 0.073 & 0.098 & 0.157 & 4.611 & 0.073 & 3.162 & 0.157\\
148 & 0.374289368 & 0.4267925092 & 0.343 & 0.006 & 0.097 & 0.026 & 0.054 & 0.007 & 0.283 & 0.075 & 0.157 & 0.021 & 0.872 & 0.004 & 0.306 & 0.218 & 0.692 & 0.039 & 5.103 & 0.218 & 3.621 & 0.039\\
160 & 0.569254354 & 0.244693639 & 0.338 & 0.012 & 0.094 & 0.012 & 0.071 & 0.017 & 0.277 & 0.035 & 0.21 & 0.05 & 0.796 & 0.004 & 0.012 & 0.013 & 0.063 & 0.172 & 4.208 & 0.014 & 1.102 & 0.172\\
163 & 0.590452294 & 0.2288151858 & 0.238 & 0.01 & 0.036 & 0.01 & 0.05 & 0.013 & 0.149 & 0.041 & 0.207 & 0.054 & 0.192 & 0.008 & 0.844 & 0.166 & 0.3 & 0.123 & 4.466 & 0.166 & 1.416 & 0.123\\
171 & 0.643185732 & 0.1916635982 & 0.171 & 0.006 & 0.025 & 0.006 & 0.042 & 0.007 & 0.146 & 0.036 & 0.245 & 0.041 & 0.239 & 0.007 & 0.664 & 0.051 & 0.277 & 0.029 & 2.74 & 0.051 & 0.378 & 0.03\\
\hline
\end{tabular}
}
\end{landscape}
\clearpage




\end{document}